\pdfoutput=1
\documentclass[11pt,nofootinbib]{revtex4-1}
\usepackage[english]{babel}
\usepackage{bbm}
\usepackage{tikz,pstricks}
\pgfarrowsdeclaredouble{feather}{feather}{to reversed}{to reversed}
\newcommand{\makeSymbol}[1]{\mathord{\vcenter{\hbox{#1}}}}

\usepackage{amsmath,amssymb,amsthm,bbold}
\numberwithin{equation}{section} 

\newcommand{\group}[2]{\mathrm{#1}\left(#2\right)}

\newcommand{\ket}[1]{|#1\rangle}
\newcommand{\bra}[1]{\langle #1|}
\newcommand{\scal}[2]{\langle #1\rangle_{#2}}
\newcommand{\intd}[1]{\mathrm{d}#1}
\newcommand{\dtensor}[3]{\begin{pmatrix}  \multicolumn{2}{c}{#1}\\#2&#3\end{pmatrix}}
\newcommand{\threej}[6]{\begin{pmatrix}  #1&#3&#5\\#2&#4&#6\end{pmatrix}}
\newcommand{\sixj}[6]{\begin{Bmatrix}  #1&#3&#5\\#2&#4&#6\end{Bmatrix}}
\newcommand{\tinysixj}[6]{\{\begin{smallmatrix}  #1&#3&#5\\#2&#4&#6\end{smallmatrix}\}}
\newcommand{\ninej}[9]{\begin{Bmatrix}  #1&#2&#3\\#4&#5&#6\\#7&#8&#9\end{Bmatrix}}
\newcommand{\tinyninej}[9]{\left\{\begin{smallmatrix}  #1&#2&#3\\#4&#5&#6\\#7&#8&#9\end{smallmatrix}\right\}}

\newcommand{\vgraph}{\mathfrak{n}}

\def\beq{\begin{equation}}
\def\eeq{\end{equation}}

\begin{document}
\title{Linking covariant and canonical LQG: new solutions to the Euclidean Scalar Constraint}
\author{Emanuele Alesci}
\email{alesci@theorie3.physik.uni-erlangen.de}
\affiliation{Universit\"at Erlangen, Institut f\"ur Theoretische Physik III, Lehrstuhl f\"ur Quantengravitation\\ Staudtstrasse 7, D-91058 Erlangen, EU}

\author{Thomas Thiemann}
\email{thiemann@theorie3.physik.uni-erlangen.de}
\affiliation{Universit\"at Erlangen, Institut f\"ur Theoretische Physik III, Lehrstuhl f\"ur Quantengravitation\\ Staudtstrasse 7, D-91058 Erlangen, EU}

\author{Antonia Zipfel}
\email{zipfel@theorie3.physik.uni-erlangen.de}
\affiliation{Universit\"at Erlangen, Institut f\"ur Theoretische Physik III, Lehrstuhl f\"ur Quantengravitation\\ Staudtstrasse 7, D-91058 Erlangen, EU}

\begin{abstract}
\begin{center}
{\bf Abstract}
 It is often emphasized that spin-foam models could realize a projection on the physical Hilbert space of canonical Loop Quantum Gravity (LQG). As a first test we analyze the one-vertex expansion of a simple Euclidean spin-foam. We find that for fixed Barbero-Immirzi parameter $\gamma=1$ the one vertex-amplitude in the KKL \cite{Kaminski:2009fm} prescription annihilates the Euclidean Hamiltonian constraint of LQG \cite{Thiemann96a}.  Since for $\gamma=1$ the Lorentzian part of the Hamiltonian constraint does not contribute this gives rise to new solutions of  the Euclidean theory. Furthermore, we find that the new states only depend on the diagonal matrix elements of the volume. This seems to be a generic property when applying the spin-foam  projector. 
\end{center}
\end{abstract}
\maketitle

\setcounter{page}{1}
 \tableofcontents
\section{Introduction}
\subsection{Motivation}
One mayor problem when quantizing Gravity is the constrained algebra, which completely determines the theory, and background independence. Canonical Loop Quantum Gravity \cite{lqgcan1,lqgcan2,lqgcan3} follows the ideas of Dirac \cite{DiracQM1964} for quantizing constrained systems and preserves background independence. The kinematical Hilbert space $\mathcal{H}_{kin}$ of LQG is spanned by spin-network functions living on semi-analytic closed graphs embedded in a three dimensional spatial hyper-surface $\Sigma$ of a 4-dimensional manifold $\mathcal{M}$. Diffeomorphism and gauge constraints can be embedded via a group averaging procedure.  The remaining constraint (Hamiltonian) is more complicated. Even though a quantization of the latter has been found \cite{Thiemann96a,Thiemann96b}, the structure of the physical Hilbert space $\mathcal{H}_{phys}$ is not fully understood up till now.\\
To circumvent the problems of the canonical theory Reisenberger and Rovelli \cite{ReisenbergerRovelli97} introduced a covariant formulation of Quantum Gravity, the so-called spin-foam model \cite{lqgcov,lqg}. This model is mainly based on the observation that the Holst action for GR \cite{HolstAction} defines a constrained BF-theory.  The strategy is first to quantize discrete BF-theory and then to implement the so called simplicity constraints. The main building block of the model is a linear two-complex $\kappa$ embedded into 4-dimensional space-time $\mathcal{M}$ whose boundary is given by an initial and final (gauge invariant) spin-network, $\psi_i$ respectively $\psi_f$, living on the initial respectively final spatial hyper surface of a foliation of $\mathcal{M}$. The physical information is encoded in the spin-foam amplitude
\beq\label{eqn:amplitudekappa}
Z[\kappa]=\prod_f\mathcal{A}_f\prod_e\mathcal{A}_e\prod_v\mathcal{A}_v\times\mathcal{B}
\eeq
where $\mathcal{A}_f\;,\mathcal{A}_e$ and $\mathcal{A}_v$ are the amplitudes associated to the internal faces, edges and vertices\footnote{In the following we will call edges and vertices in the boundary links respectively nodes to distinguish between the two-complex and the graph.} of $\kappa$ and $\mathcal{B}$ contains the boundary amplitudes.   Each spin-foam can be thought of as generalized Feynman diagram contributing to the transition amplitude from an ingoing spin-network to an outgoing spin-network. By summing over all possible two-complexes one obtains the complete "transition amplitude" between $\psi_i$ and $\psi_f$.\\
Unfortunately the simplicity constraint is second-class and the procedure how to implement it is still under debate \cite{Alexandrov:2010un}. Nevertheless substantial progress has been achieved during the last years \cite{EarlySpinfoams}. Especially, the introduction of a new vertex amplitude by Engle, Pereira, Rovelli and Livine and independently  by Freidel and Krasnov \cite{Engle:2007uq} and the introduction of an abstract model \cite{Kaminski:2009fm} led to a major breakthrough.\\
Instead of considering spin-foams as a "sum over histories" one could equally well think of spin-foams as some group averaging procedure to implement the Hamiltonian constraint in the canonical formalism (see \cite{SFscalarproduct,ReisenbergerRovelli97}). Suppose we have a family of first-class constraints $(\hat{C}_I)_{I\in\mathcal{I}}$ which form a Lie-algebra. Generically, the point zero does not lie in the point spectrum of the constraint operators and therefore the eigenvectors can not form the entire solution space. To obviate this problem one has to consider generalized eigenvectors $l\in\mathcal{D}^{\ast}_{\text{kin}}$ in the algebraic dual of a dense domain of $\mathcal{H}_{\text{kin}}$ such that
\beq
\left[(\hat{C}_I)'l\right](\psi):=l(\hat{C}_I^{\dagger}\psi)=0 \qquad\forall\, I\in\mathcal{I},\, \psi\in\mathcal{D}_{\text{kin}}
\eeq
where $(\hat{C}_I)'$ is the dual operator on $\mathcal{D}^{\ast}_{\text{kin}}$. The space of generalized solutions $\mathcal{D}^{\ast}_{\text{phys}}$ is a proper subspace of $\mathcal{D}^{\ast}_{\text{kin}}$.  In order to construct a physical Hilbert space one considers $\mathcal{D}^{\ast}_{\text{phys}}$ as the algebraic dual of a dense subspace $\mathcal{D}_{\text{phys}}\subset\mathcal{H}_{\text{phys}}$ so that all observables are densely defined in $\mathcal{H}_{\text{phys}}$. The inner product on $\mathcal{H}_{\text{phys}}$ is chosen such that adjoints in the physical scalar product represent adjoints in the kinematical one. It can be systematically constructed by an anti-linear map, called rigging map,
\beq
\eta:\mathcal{D}_{\text{kin}}\to\mathcal{D}^{\ast}_{\text{kin}}
\eeq
such that
\beq
\scal{\eta[\phi]|\eta[\psi]}{\text{phys}}:=\eta[\phi](\psi)\qquad \phi,\psi\in\mathcal{D}_{\text{kin}}
\eeq
and
\beq
\hat{O}'\eta[\phi]=\eta[\hat{O}\phi]\qquad\forall \phi\in\mathcal{D}_{\text{kin}}\,.
\eeq
The physical Hilbert space is subsequently defined by the completion of  $\mathcal{D}_{\text{phys}}:=\eta(\mathcal{D}_{\text{kin}})\setminus\text{ker}(\eta)$.\footnote{For more details on the construction of a rigging map see e.g. \cite{lqgcan2}.} Strictly speaking, such a construction only works for closed, first-class constraints. But the constraint algebra in GR is open with structure functions instead of structure constants. Nevertheless, it is often emphasized that spin-foams could provide such a rigging map even though one starts with a different action and constraint algebra than in the canonical approach. If this is indeed the case then the physical inner product would be given by
\beq
\label{eqn:physscalr}
\scal{\phi|\psi}{\text{phys}}=\sum_{\kappa:\psi\to\phi}Z[\kappa]
\eeq
and the rigging map would correspond (schematically) to
\beq
\label{eqn:project}
\eta[\psi]=\sum_{\phi\in\mathcal{H}_{\text{kin}}}\;\sum_{\kappa:\psi\to\phi}Z[\kappa]\bra{\phi}\;.
\eeq
Since all constraints are satisfied in $\mathcal{H}_{\text{phys}}$ the so-defined physical scalar product must obey
\beq
\label{eqn:constraint-proj}
\scal{\psi_{out}|\hat{C}^{\dagger}|\psi_{in}}{\text{phys}}=\sum_{\phi\in\mathcal{H}_{\text{kin}}}\;\sum_{\kappa:\psi_{out}\to\phi}Z[\kappa]\scal{\phi|\hat{C}^{\dagger}|\psi_{in}}{\text{kin}}=0
\eeq
for all $\psi_{out},\,\psi_{in}\in\mathcal{H}_{\text{kin}}$. This is clearly the case for the gauss constraint because the boundary of a spin-foam are gauge invariant spin-networks. The diffeomorphism constraint is harder to deal with since spin-foams are defined on a discretization of space-time and break diffeomorphism invariance. But in the abstract formulation the amplitudes do not depend on the embedding and one can implement the constraint by restricting on equivalence classes of spin-networks. Wether the Hamiltonian constraint also obeys \eqref{eqn:constraint-proj} depends crucially on the definition of the vertex amplitudes. However, it is fcrucially to solidify LQG results that compute transition amplitudes assuming the EPRL-FK model as defining the dynamics of the theory in the context of the propagator \cite{scattering1,scattering3, Bianchi:2010zs}
\subsection{Outline}
As a first test for \eqref{eqn:constraint-proj} we consider an easy spin-foam amplitude and show that 
\beq
\label{eqn:onev}
\sum_{\phi} \mathcal{Z}[\kappa] \scal{\phi|\hat{H}_{\vgraph}|\psi_{in}}{}=0
\eeq
where $\kappa$ is a two-complex with only one internal vertex such that $\phi$ is a spin-network induced on the  boundary of $\kappa$ and $\hat{H}_{\vgraph}$ is the Hamiltonian constraint acting on the node $\vgraph$.\\
In Sec. II. we briefly review the quantization of the Hamiltonian constraint \cite{Thiemann96a} and compute the action on three- and four-valent nodes by employing graphical calculus. Recall, that the full constraint $\mathcal{C}=-[H+(s-\gamma^2)H_L]$ can be decomposed into its Lorentzian and Euclidean part, $H_L$ respectively $H$ where $\gamma$ is the Barbero-Immirzi parameter and $s$ is the signature of the metric. We restrict the analysis to the Euclidean sector with $s=1,\gamma=1$. Then $\mathcal{C}$ reduces to the Euclidean part only. The operator $\hat{H}$ acts locally on nodes  and is graph-changing. We choose a tetrahedral regularization of the latter as proposed in \cite{Thiemann96a}. Then roughly speaking, $H$ creates a new link connecting two pairwise distinct links adjacent to the same node. In order to keep the calculation as simple as possible we choose $\psi_{out}=\psi_{in}$ being a spin-network with two $n$-valent nodes.\\
We will summarize the construction of the spin-foam amplitude in Sec.III. \\
In the subsequent section we evaluate the spin-foam amplitude for a two-complex $\kappa$ with only one internal vertex and boundary $\partial\kappa=\psi_{out}\cup\phi$ such that $\kappa$ is a tube $\psi_{out}\times[0,1]$ with an additional face between the internal vertex and the new link created by $\hat{H}$ (see Fig.\ref{fig:foam}). \\
In Sec. IV  we show that the one-vertex amplitude annihilates the Hamiltonian constraint by employing basic summation identities of 6j-symbols, when acting on three respectively four-valent nodes. For an n-valent node the sum \eqref{eqn:onev} is a sum over spin-networks based on $\binom{n}{2}$ different graphs. Remarkably, each partial sum over spin-networks based on the same graph vanishes. This shows, that the solutions constructed via the spin-foam method build a proper subset of $\mathcal{H}_{\text{phys}}$. As an important side result we find that $\sum_{\phi}\;Z[\kappa]\bra{\phi}$ selects only those matrix elements of $\hat{H}$ which depend on the diagonal matrix elements of the volume. \\
Sec. V contains a summary of our results and give an outlook to open questions.
\section{Hamiltonian constraint}
\label{sec:Ham}
A primary quantum version of the Hamiltonian constraint operator was introduced by Rovelli and Smolin \cite{loops}.  The operator proofed to act only on the nodes of a spin-network function. But it was divergent on general states. Latter \cite{HamiltonianRS}, it was shown that the following two properties are crucial in order to obtain a well defined finite Hamiltonian operator in the background independent context:
\begin {itemize}
\item the operator needs to be a density (more precisely, a three-form)
\item diffeomorphism invariance trivializes the limit when the regulator is removed from the operator. 
\end{itemize}
The first requirement forces us to use a non-polynomial version of the constraint which quantization is much more involved. After many efforts \cite{Early-hamiltonians, Late-hamiltonians,HamiltonianR}, it was suggested \cite{Thiemann96a,Thiemann96b} to express the inverse triad $e$ as the Poisson bracket between the volume $V$ and the holonomy  $h$ of the Ashtekar connection $A$: $ e\sim h^{-1}[h,V]$. This trick made it possible to construct an Hamiltonian with the above properties which can be regularized on a given triangulation $T$ of the space manifold. (For criticism see \cite{dubbi}.)\\
In the following two sections we will review the basic construction of $\mathcal{C}$ as proposed in \cite{Thiemann96a} in order to clarify the model and our notation. The reader familiar to the framework can easily skip the next two sections.  
\subsection{Hamiltonian constraint}
\begin{figure}
  \begin{center}

\includegraphics{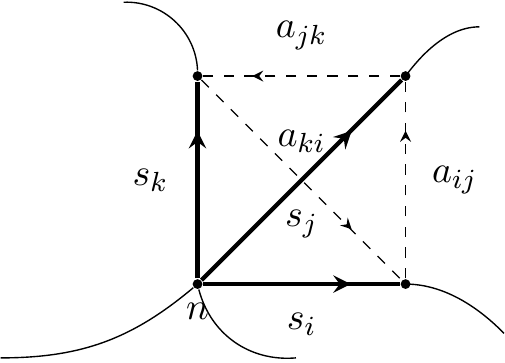}
   \parbox{9cm}{\caption[]{\label{tetrahedron} \small
       An elementary tetrahedron $\Delta \in T$ constructed
      by adapting it to a graph $\Gamma$ which
       underlies a cylindrical function.}}
  \end{center}
\end{figure}

The classical Hamiltonian constraint is 
\beq
{\cal C}=-\frac{2}{\kappa}{\rm Tr}[(\,^{(\gamma)}F-(\gamma^2-s)K\wedge K) \wedge e]
\label{constraint}
\eeq
where $e=e^i_a \tau_i dx^a$ is the inverse triad, $K $ the extrinsic curvature, $\,^{(\gamma)}F$ the curvature of the Ashtekar connection with real Immirzi parameter $\gamma$ and $s$ the signature. In the following we choose units such that $\kappa/2:=\frac{8\pi G}{c^3}=1$. \\
The constraint can be split into its  ``Euclidean"  part  $H={\rm Tr}[F  \wedge e]$ and Lorentzian part $H_L= C- H$.  Following \cite{Thiemann96a}, we can rewrite \eqref{constraint} by using
\begin{gather}
\begin{gathered}
  \label{keyID1}
e^i_a =-2 \{A_a^i(x),V\} \\
K^i_a=2\{A^i_a(x),K\}\\
K=2\{ H,V\}
\end{gathered}
\end{gather}
where  $V$ is the volume of an arbitrary region $\Sigma$ containing the point $x$. Smearing the constraints with lapse function $N(x)$ gives
\begin{align}
\begin{split}
  H[N] =&   
       \int_\Sigma d^3x \, N(x)\,H(x) \\
     =&  - 2 \int_\Sigma  \, N  \ {\rm Tr}(F \wedge   \{ A, V \})             
  \label{H_E_N}  
 \end{split}
 \\
 \begin{split}
  H_L[N] =&   
       \int_\Sigma d^3x \, N(x)\, H_L(x) \\
     = -(\gamma^2&-s)  \int_\Sigma  \, N  \ {\rm Tr}(\{A,\{ H_L,V\}\} \wedge \{A,\{H_L,V\}\} \wedge \{ A, V \})             \,.
\end{split}
\end{align}
This expression requires a regularization in order to obtain a well-defined operator on $\mathcal{H}_{kin}$. Up to now there exist many different proposals (see e.g \cite{lqgcan1,EmanueleReg}). We will follow the original proposal \cite{Thiemann96a} and use a triangulation $T$  of the manifold $\Sigma$ into elementary tetrahedra with analytic links adapted to the graph $\Gamma$ of an arbitrary spin-network.  For each pair of links $e_i$ and $e_j$ incident at a node $\vgraph$ of $\Gamma$ we choose semi-analytic arcs $a_{ij}$ such that the end points $s_{e_i},s_{e_j}$ are interior points of $e_i$ respectively $e_j$ and  $a_{ij}\cap\Gamma=\{s_{e_i},s_{e_j}\}$. The arc $s_i$ is the segment of $e_i$  from $\vgraph$ to $s_i$ and $s_{i}$, $s_{j}$ and $a_{ij}$ generate a triangle $\alpha_{ij} := s_{i} \circ a_{ij} \circ s_j^{-1}$. Three (non-planar) links define a tetrahedron (see Fig. \ref{tetrahedron}). Now we can decompose \eqref{H_E_N} into a sum of one term per each tetrahedron of the triangulation
\begin{eqnarray}
  \label{Ham_T1}
  H[N] = \sum_{\Delta \in T}
        {-2}\, \int_\Delta d^3x \, N 
            \ \epsilon^{abc}\ {\rm Tr}(F_{ab}
            \{ A_c, V \}) ~.
          \label{Ham_T2}
\end{eqnarray}
Define the classical {\em regularized\/} Hamiltonian constraint
\beq
   H_{T}[N]:=  \sum_{\Delta \in T} H_{\Delta} [N] ~.
        \label{H_delta:class_equiv}
  \eeq 
The connection $A$ respectively the curvature are regularized as usual by the holonomy $h_{s}:=h[s]\in\group{SU}{2}$ (in the fundamental representation $m=1/2$) along the segments $s_i$ respectively along the loop $\alpha_{ij}$. This yields
\begin{align}
\begin{split}
      H_{\Delta}[N]  :&= - \frac{2}{3} \, N(\vgraph) \, \epsilon^{ijk} \, \mathrm{Tr}\Big[h_{\alpha_{ij}} 
     	h_{s_{k}} \big\{h^{-1}_{s_{k}},V\big\}\Big]
 \label{H_delta}
\end{split}
\end{align}
and converges to the Hamiltonian constraint \eqref{Ham_T2} if the triangulation is sufficiently fine. The expression (\ref{H_delta:class_equiv}) can finally  be promoted to a quantum operator,  since volume and holonomy have corresponding well-defined operators in LQG.  The lattice spacing of the triangulation $T$ that acts as a regularization parameter  can be removed in a suitable operator topology, see  \cite{Thiemann96a} for details. \\
\mbox{}\\
{\bf Remarks}
\begin{itemize}
\item In  \cite{Gaul:2000ba} it was pointed out that the operator can be immediately generalized by replacing the trace in \eqref{Ham_T2} with a trace in an arbitrary irreducible representation $m$: $\mathrm{Tr}_{m}[U]=\mathrm{Tr}[R^{(m)}(U)]$ where $R^{(m)}$ is a matrix representation of $U\in\group{SU}{2}$. Equation \eqref{H_delta} can thus be replaced by
 \beq
   \label{Hm_delta:classical2}
   H^m_{\Delta}[N]:= \frac{N(\vgraph)}{N^2_m} \,  \, \epsilon^{ijk} \,
   \mathrm{Tr}\Big[h^{(m)}_{\alpha_{ij}} h^{(m)}_{s_{k}} \big\{h^{(m)-1}_{s_{k}},V\big\}\Big] ~,
 \eeq
 $N_m^2=\mathrm{Tr}_m[\tau^i \tau^i]=-(2m+1)m(m+1)$ and $h^{(m)}=R^{(m)}(h)$.  As shown in \cite{Gaul:2000ba}, this converges to $H[N]$ as well.\\
\item The Lorentzian part of the constraint can be regularized by a similar method.
\end{itemize}
 \subsection{Properties}
In this section we will summarize the important properties of the Euclidean Hamiltonian constraint.\\
It is immediate to see that when acting on a spin-network state, the operator reduces to a sum over terms each acting on individual nodes. Acting on nodes of valence $n$ the operator gives
\beq
   {\hat{H}}^m_{\Gamma}[N] \, \psi_\Gamma = \frac{i}{\hbar}
  \sum_{\vgraph \in \mathcal{N}(\Gamma)}
  \sum_{\vgraph(\Delta) = \vgraph}  \frac{p_{\Delta}}{E(\vgraph)} {\hat{H}}^m_{\Delta} [N]
\, \psi_\Gamma ~,
\eeq
where ${H}^m_{\Delta}$ is the quantum version of \eqref{Hm_delta:classical2}, $\mathcal{N}(\Gamma)$ is the set of nodes of $\Gamma$ and $E(\vgraph) =\binom{n}{3}$ is the number of unordered triples of links adjacent to $\vgraph$.  The second sum is a sum over tetrahedra with a node at $\vgraph$ and not intersecting with other nodes of $\Gamma$. Moreover, $p_{\Delta}=1$, whenever $\Delta$ is a tetrahedron having three edges coinciding with three links of the spin-network state, that meet at the node $\vgraph$, otherwise $p_{\Delta}=0$.\\
On diffeo\-mor\-phism invariant states $\phi\in\mathcal{H}_{phys}\subset\mathcal{H}^{\ast}_{kin}$ the regulator dependence drops out trivially because two operators $\hat{H}$ and $\hat{H}'$ that are related by a refinement of the triangulation differ only in the size of the loops $\alpha_{ij}$. Therefore, the resulting states are in the same equivalence class and
\beq
\label{eqn:dual_act}
[\hat{H}^{\dagger}\phi](\psi):=\langle \phi ,\hat{H}\psi \rangle = \langle \phi, \hat{H}' \psi \rangle~,
\eeq 
in $\mathcal{H}_{diff}$. This proves that the Hamiltonian constraint on diffeomorphism invariant states is independent from the refinement of the triangulation.\\
The action $\hat{H}(N)$ on a spin-network state  $T_{\Gamma,\vec{j},\vec{c}}$ defined on a graph $\Gamma$ results in a finite linear combination of spin-network states defined on graphs $\Gamma_I$ where $\Gamma\subset\Gamma_I$ and $a_I:=\Gamma_I-\Gamma$ is  produced by one of the arcs $a_{ij}(\Delta)$, which carries spin $j_I=m$. The new nodes are called \emph{extra ordinary}. In some cases it can happen that links connecting the original node with the new extraordinary nodes carry trivial representation if this is allowed by the recoupling conditions.  Extraordinary nodes are at most trivalent and intersections of precisely two analytic curves $c,c'\subset\Gamma$, that is, $\vgraph=c\cap c'$, such that $\vgraph$ is an endpoint of $c$ but not of $c'$.  A link $e$ of a graph $\Gamma$ is called extraordinary provided that its endpoints $\vgraph_1,\vgraph_2$ are both extraordinary nodes. Furthermore, those links are adversed to a node $\vgraph$ of $\gamma$ which is incident to at least three links $s_1,s_2,s_3$ with linearly independent tangents at $\vgraph$ such that  $s_1$/$s_2$ connect $\vgraph$ and $\vgraph_1$/$\vgraph_2$. We will call $\vgraph$ the typical node associated with $e$. All links produced by $\hat{H}$ are extraordinary. Since the volume operator annihilates coplanar nodes and gauge invariant nodes of valence three (only true for Ashtekar-Lewandowski version, see \cite{lqgcan1}) $H$ does not act on extraordinary nodes.
\subsection{Action on a trivalent node}
Let us now compute the action of the operator ${\mathcal{H}}^m_{\Delta}$ on a trivalent node where all links are outgoing,  following \cite{Gaul:2000ba,EmanueleReg}.  Denote a trivalent node by  
$\ket{\vgraph(j_i,j_j,j_k)} \equiv \ket{\vgraph_3}$, whereas $j_i,j_j,j_k$ are the
spins of the adjacent links $e_i,\, e_j,\, e_k$:
\beq
\left|\vgraph_3\right\rangle=\makeSymbol{\includegraphics{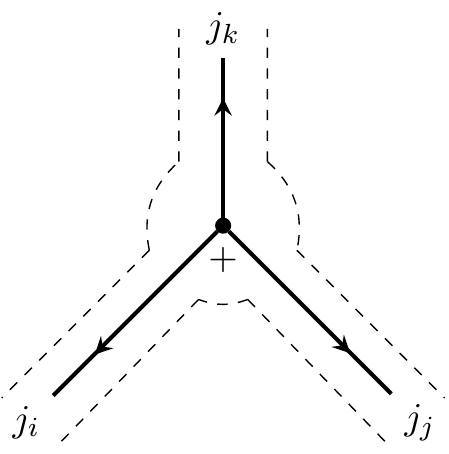}}
\eeq
Note, the links are also labeled by group elements with orientations indicated by the arrows. In order to simplify the graphics we only displayed the node and its adjacencies. Furthermore, everything contained in the dashed circle belongs to the node and everything between the dashed lines belong to the same link. \\
When quantizing expression \eqref{Hm_delta:classical2} the holonomies and the volume are replaced by their corresponding operators and the Poisson bracket is replaced by a commutator. Since the volume operator vanishes on a gauge invariant trivalent node we only need to compute 
\begin{equation}
  \label{H_m_Delta} 
    \hat{H}^m_{\Delta} \, \ket{\vgraph_3} =N_{\vgraph} \, 
  \epsilon^{ijk} \, \mathrm{Tr} \left(
  \frac{\hat{h}^{(m)}[\alpha_{ij}] - \hat{h}^{(m)}[\alpha_{ji}]}{2} \,
  \hat{h}^{(m)}[s_{k}]\, \hat{V} \,  
  \hat{h}^{(m)}[s^{-1}_{k}] \right) \ket{\vgraph_3} ~,
\end{equation}
where all (global) constants have been absorbed in the lapse function $N_{\vgraph}$.  Antonia The operator $ \hat{h}^{(m)}[s^{-1}_{k}]$, corresponding to the holonomy along a segment $s_k$ with reversed orientation, acts by multiplication with $R^{(m)}(h_{s_k^{-1}})$ along $s_k$. The matrix $R^{(m)}$ can be recoupled using \eqref{eqn:basicrule} and \eqref{eqn:invertG}. Thus, $ {h}^{(m)}[s^{-1}_{k}]$ creates  a free index in the $m$-representation located at the node (inside the dashed circle), making it non-gauge invariant and a new node on the link $e_k$:
\beq
  \label{holonomy_action1} 
  \hat{h}^{(m)}[s^{-1}_{k}]~\left|\vgraph_3\right\rangle
      ~= (-1)^{2m}\sum_c \, d_c\!\!\!
      \makeSymbol{\includegraphics{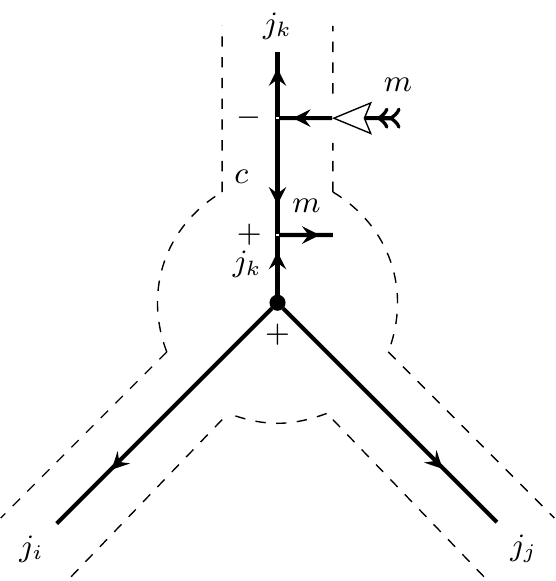}}
\eeq
where $d_c=2c+1$ is the dimension of $c$. The range of the sum over the spin $c$ is determined by the Clebsch-Gordan conditions and the little flag represents the group element $h^{-1}_{s_k}$. \\
The volume operator now acts on a trivalent non-gauge invariant node with a virtual link in $j_k$ representation. The matrix elements of the volume operator \cite{RovelliSmolin95,AshtekarLewand98,Lewandowski96}, have been computed in  \cite{DePietriRovelli96,Brunnemann:2004xi} and the results have been applied to the Hamiltonian constraint operator in \cite{Borissovetal97,Gaul:2000ba}.
\\
The operators $ {h}^{(m)}[\alpha_{ij}]\,  {h}^{(m)}[s_{k}]$ and $ {h}^{(m)}[\alpha_{ji}]\,  {h}^{(m)}[s_{k}]$ add open loops with opposite orientations $\alpha_{ij}$ and $\alpha_{ji}$, where we fix $\alpha_{ij}$ to be oriented anti-clockwise. Like above, the representations living on the same link can be recoupled. The $m$-trace connects the free ends of the open loops to the two open links in $ {h}^{(m)}[s^{-1}_{k}]\, \ket{\vgraph}$ taking into account the orientations. Finally one has to use \eqref{eqn:basicrule2}:
\begin{gather}
\begin{gathered}
 \label{trace_part}
  \mathrm{Tr}\left(
  \frac{ {h}^{(m)}[\alpha_{ij}] -  {h}^{(m)}[\alpha_{ji}]}{2} \:
   {h}^{(m)}[s_{k}]\,  {V} \,  {h}^{(m)}[s^{-1}_{k}] 
  \right) \, |\vgraph (j_i,j_j,j_k) \rangle \\
  =\sum_{a,b} 
        \, A^{(m)}(j_i,a|j_j,b|j_k) 
      \makeSymbol{\includegraphics{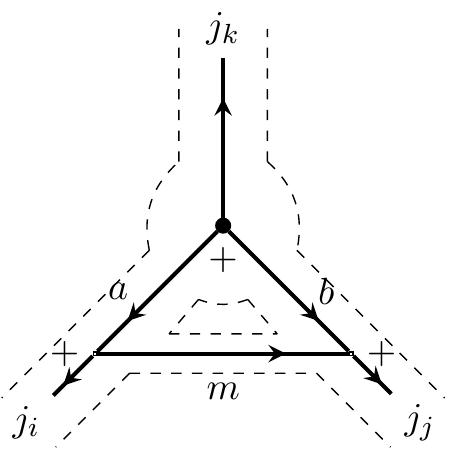}}     
\end{gathered}
\end{gather}
where the range of the sums over $a,b$ is determined by the  Clebsch-Gordan conditions\footnote{The link $m$ can not be removed with \eqref{eqn:basicrule2} since this is a pure recoupling identity but $m$ also carries a group element.} and
\begin{align}
   \label{amplitude}
 &A^{(m)}(j_i,a|j_j,b|j_k) := \sum_c\; \lambda^{mj_i}_a 
        \lambda^{mj_j}_b  \lambda^{mj_k}_c \;(-)^{j_i+j_j-j_k}\;d_a d_b d_c\nonumber\\
    &\times\;\sum_{\beta(j_i,j_j,m,c)}  \!\!\!\! V{}_{j_k}{}^\beta (j_i,j_j,m,c) \\
&\times\;\left[ \lambda_c^{m\beta}(-)^{a+j_j+c}\; \tinysixj{a}{\beta}{j_j}{m}{c}{j_i}\tinysixj{a}{m}{b}{c}{j_k}{j_j}                 
  - 
   \lambda^{mj_k}_c (-)^{b+j_i+c}\tinysixj{j_i}{m}{b}{\beta}{c}{j_j}\tinysixj{a}{c}{b}{m}{j_k}{j_i}
   \right] ~.
   \nonumber
\end{align}
The sign factors are due to the chosen orientation which has to be respected when applying the recoupling identities (see \ref{graphical}) and can be manipulated by realizing that $(-)^{2a+2b+2c}=1$ if $a,b,c$ fulfill the Clebsch-Gordan conditions. The summation index $\beta = \beta(j_i,j_j,m,c)$ which appears due to the non-diagonal action of the volume operator, ranges on the values which are determined by the simultaneous admissibility of the trivalent nodes $\{j_i,j_j,\beta \}$ and $\{m,c,\beta\}$. If $m=\frac{1}{2}$ then the volume operator acts diagonally and $\beta=j_k$.\footnote{We get a correction of sign factors compared to \cite{EmanueleReg}. This correction is necessary in order that the action of $\mathrm{Tr}(F_{ij})$ vanishes.}\\
The complete action of the operator on a trivalent state  $|\vgraph (j_i,j_j,j_k) \rangle$ can be obtained by contracting the trace part (\ref{trace_part}) with  $\epsilon^{ijk}$. Thus, $\hat{H}$ projects on a linear combination of three spin-networks which differ by exactly one new link labeled by $m$ between each couple of the 'old' links at the node. 
\subsection{Action on a 4-valent node}
  The computation for a 4-valent node $\ket{\vgraph_4}$ is similar to the previous: 
\begin{equation}
  \label{H_m_Delta1} 
    \hat{H}^m_{\Delta} \, \ket{\vgraph_4} = N_{\vgraph} \, 
  \epsilon^{ijk} \, \mathrm{Tr} \left(
  \frac{\hat{h}^{(m)}[\alpha_{ij}] - \hat{h}^{(m)}[\alpha_{ji}]}{2} \,
  \hat{h}^{(m)}[s_{k}]\, [\hat{V},  \hat{h}^{(m)}\left[s^{-1}_{k}]\right] \right) \ket{\vgraph_4} ~.
\end{equation}
In the subsequent calculation we fix choose all links to be outgoing from the node
\beq
\ket{\vgraph_4}=\sqrt{d_i}
\makeSymbol{\includegraphics{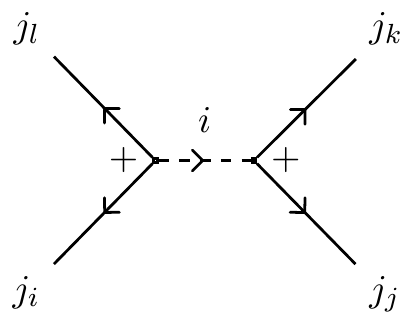}}
\eeq
where $i$ labels the intertwiner (inner link). Furthermore, we fix the orientation of the loop $\alpha_{ij}$ in \eqref{H_m_Delta1} to be anti-clockwise.  The part $\mathrm{Tr} \left(
  \hat{h}^{(m)}[\alpha_{ij}] - \hat{h}^{(m)}[\alpha_{ji}] \, \hat{V} \right) \ket{\vgraph_4}$ vanishes since the volume does not modify the representations but the trace is taken in the representation space and $\mathrm{Tr} (\hat{h}^{(m)}[\alpha_{ij}] - \hat{h}^{(m)}[\alpha_{ji}])=0$.\\
For the other part, the holonomy $ \hat{h}^{(m)}[s^{-1}_{k}]$  changes  the valency of the node and the Volume subsequently acts on the 5-valent non-gauge invariant node. Graphically this corresponds to
\beq
\hat{V}\;\hat{h}^{(m)}[s^{-1}_{k}] \; \ket{\vgraph_4}=(-1)^{2m}\sum_c d_c\sum_{\beta,\gamma}\; V_{i,j_k}^{\;\;\gamma, \beta}\!\!
\makeSymbol{\includegraphics{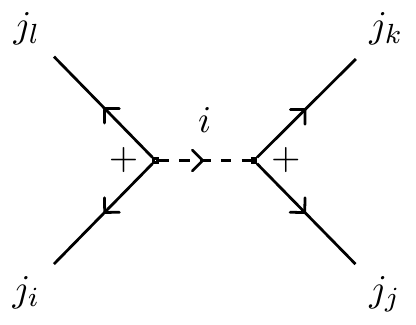}}.
\eeq 
Finally one arrives at
\beq
\begin{split}
&\mathrm{Tr} \left(
  \frac{\hat{h}^{(m)}[\alpha_{ij}] - \hat{h}^{(m)}[\alpha_{ji}]}{2} \,
  \hat{h}^{(m)}[s_{k}]\, \hat{V} \,  
  \hat{h}^{(m)}[s^{-1}_{k}] \right) \ket{\vgraph_4} =\\
  &\sum_{a,b,c}d_a d_b d_c \sum_{\beta,\gamma} V_{i,j_k}^{\;\;\gamma, \beta}\;\;
  \Big[
\makeSymbol{\includegraphics{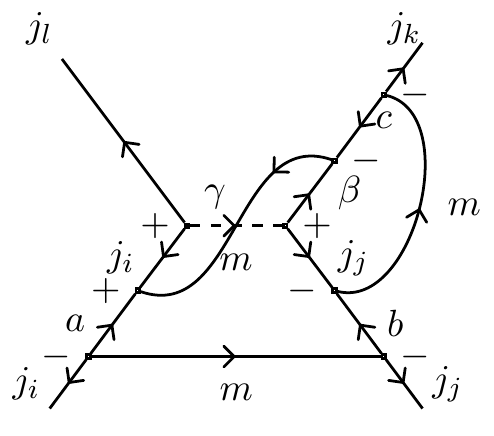}}
-
\makeSymbol{\includegraphics{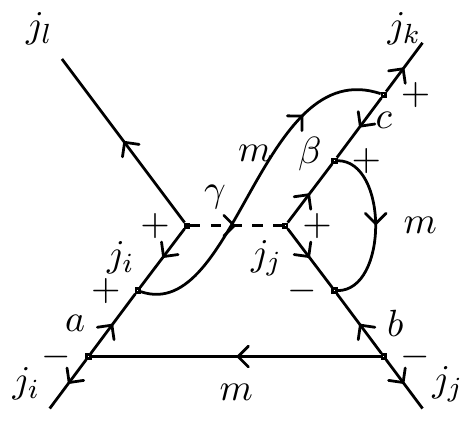}}
  \Big]
\end{split}
\eeq
This can be simplified using \eqref{ac} and \eqref{eqn:basicrule2}
\beq
\begin{split}
 &\hat{H}^m_{\Delta} \, \ket{\vgraph_4}
 = \sum_{a,b,c}d_a d_b d_c \sum_{\beta,\gamma} V_{i,j_k}^{\;\;\gamma, \beta}
 \Big[
  (-1)^{2j_i+a+j_l-j_j-\beta+m}\\
&\sum_{\alpha} d_{\alpha}
\sixj   { \gamma}  { a} {\alpha}    {j_i} {m}  {j_l} 
\sixj    {\gamma}  {c} {\alpha} {\beta} {m}  {j_j} 
\makeSymbol{\includegraphics{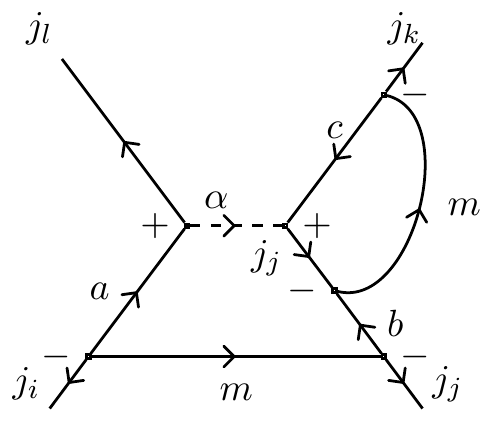}}\\
&-
(-1)^{\gamma+j_j+c+m} \left\{\begin{array}{ccc}     \gamma  & j_j & \beta \\      m & c & b                 \end{array}\right\}
 \makeSymbol{\includegraphics{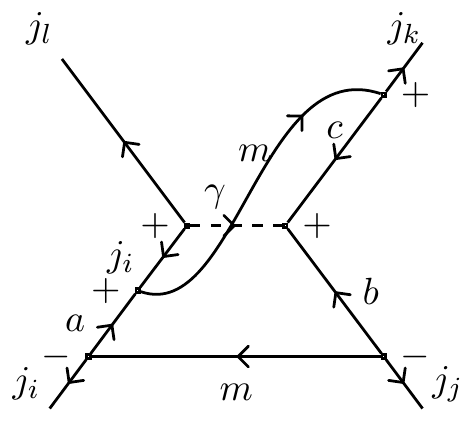}}  
			\Big]\\
  &=
\sum_{a,b,c}d_a d_b d_c \sum_{\beta,\gamma} V_{i,j_k}^{\;\;\gamma, \beta}\;\;
  (-1)^{2j_i+j_l+j_j+a+m}
\sum_{\alpha} d_{\alpha}  \Big[(-1)^{\alpha+\beta-c-j_j}
\tinysixj   { \gamma}  { a} {\alpha}    {j_i} {m}  {j_l} 
\tinysixj    {\gamma}  {c} {\alpha} {\beta} {m}  {j_j} 
\tinysixj     {\alpha} {m}   {j_j}{j_k} {c}  {b} \\
&-(-1)^{c+\gamma-b-j_k}  \tinysixj{  \gamma}   { m}  {j_j} {c} {\beta} {b}   \tinysixj{  \gamma}   { a}  {\alpha} {j_i} {m} {j_l} \tinysixj{  \gamma}   { j_k}  {\alpha} {c} {m} {b}
\Big]
\makeSymbol{\includegraphics{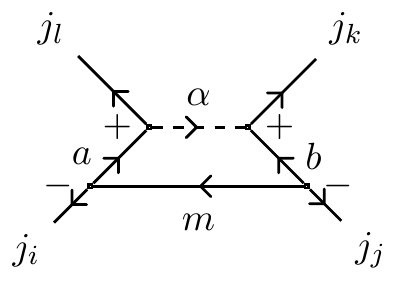}}
\end{split}
\eeq
The sign factors are due to the chosen orientation and can be manipulated as above. 
\section{Spin-foam}
\subsection{The model}
\label{ssec:model}
In this section we briefly recall the definition of Euclidean spin-foam models as suggested by Kaminski, Kisielowski and Lewandowski \cite{Kaminski:2009fm} and clarify our notation. Since we are only interested in the evaluation of a spin-foam amplitude  we choose a combinatorial definition of the model. \\
Consider an oriented two-complex $\kappa$ defined as the union of the set of faces (2-cells) $\mathcal{F}$, edges (1-cells) $\mathcal{E}$ and vertices (0-cells) $\mathcal{V}$ such that every edge $e$ is a 1-face\footnote{For a definition of complex see e.g. \cite{pl}} of at least one face $f$ (notation: $e\in\partial f$) and every vertex $v$is a 0-face of at least one edge $e$ (notation: $v\in\partial e$). We call edges which are contained in more than one face $f$ internal and denote the set of all internal edges by $\mathcal{E}_{int}$. Vice versa all vertices adjacent to more than one internal edge are also called internal and denote the set of these vertices by $\mathcal{V}_{int}$. The boundary $\partial\kappa$ is the union of all external vertices (called: nodes) $\vgraph\notin\mathcal{V}_{int}$ and external edges (called: links) $l\notin\mathcal{E}_{int}$. If $\partial\kappa$ forms a closed but possibly disconnected graph and the orientation of $e\in\partial\kappa$ agrees with the orientation induced by the unique face $f, e\in\partial f$ (we say: $f$ is ingoing to $e$) then $\kappa$ is called a proper foam. In the following we will only consider proper foams.\\
A spin-foam is a triple $(\kappa,\rho_f,I_e)$ consisting of a proper foam whose faces are labeled by irreducible representations of a Lie-group $G$ (here $\group{SO}{4}$) and whose internal edges are labeled by intertwiners $I$. This induces a spin-network structure $\partial(\kappa,\rho_{l_f},I_{\vgraph_e})$ on the boundary of $\kappa$. In the following we will denote the pair $(v,f)$ such that $v\in\partial f$ by $v_f$ and analogously for all other pairings $e_v,e_f$ etc. Furthermore $\partial v$ is the set of all faces $f_v$ and edges $e_v$.\\
Suppose $\kappa$ is foam without boundary. Following \cite{Ding:2010fw}, we label each edge $e\in\kappa$ by an group element $e\to U_e\in{\group{SO}{4}}$ such that 
\beq
U_e= g_{e_{s(e)}}g^{-1}_{e_{t(e)}}
\eeq
where $s(e)$ / $t(e)$ is the source / target of $e$. For each pair $(v,f)$ with $v\cap f= v$ and edges $e\cap e'=v$, $e,e'\in\partial f$ we define
\beq
g_{f_v}:=(g_{e_v}^{-1}g_{{e'}_v})^{\epsilon_{e_f}}
\eeq 
where $\epsilon_{e_f}=\pm$ according to the orientations. With this definitions the BF partition function can be rewritten as 
\beq
\label{eqn:zbf}
Z^{BF}[\kappa]=\int_{\group{SO}{4}}\intd{g_{f_v}} \prod_{f\in\kappa}\delta\left(\prod_{v\in\partial f}g_{f_v}\right)\prod_{f_v} \underbrace{\int\intd{g_{e_v}}\delta(g_{f_v}^{-1}g_{e_v}g_{{e'}_v}^{-1})}_{\mathcal{A}_v(g_{f_v})}
\eeq
Note, $\mathcal{A}_v(g_{fv})$ defines an $\group{SO}{4}$ invariant function on the graph $\Gamma_v$ induced on the boundary of the vertex $v$ \cite{Ding:2010fw}. As it is well known, the boundary Hilbert space $\mathcal{H}_v$ is spanned by (normalized) spin-network functions $T^{BF}_{\Gamma_v,\rho,I}(g_f)$\footnote{The links $l_f$ bounding the face $f$ are labeled by irreducible representations $\rho_f$ and nodes $n_e$ bounding the edge $e$ are labeled by intertwiners $I_e$ as usual.}
\beq
\mathcal{A}_v(g_f)=\sum_{\rho_f,I_e}\;\prod_{f\in\partial v}\sqrt{\dim{\rho_f}}\quad\mathrm{Tr}_v\left(\bigotimes_{e\in\partial v} I^{\dagger}_e\right)\quad T^{BF}_{\Gamma_v,\rho,I}(g_f)  
\eeq
Locally $\group{SO}{4}\sim\group{SU}{2}\times\group{SU}{2}$ which implies $\rho_{\group{SO}{4}}=\rho^+_{\group{SU}{2}}\otimes\rho^-_{\group{SU}{2}}$ and $T^{BF}_{\Gamma_v,\rho,I}(g_f) =T_{\Gamma_v,j^+,\iota^+}(g^+_f) \otimes T_{\Gamma_v,j^-,\iota^-}(g^-_f)$.\\
In the EPRL model \cite{Engle:2007uq} the simplicity constraint is imposed weakly. Consequentially, we have to restrict $\mathcal{A}_v(g_f)$ to the EPRL subspace $\mathcal{H}_v^{EPRL}$ spanned by the functions 
\beq 
\label{eqn:boundarysn}
\begin{split}
T^E_{\Gamma_v,j_f,\iota_e}=&\prod_{f_v}\sqrt{d_{j_{f_v}^+}\;d_{j_{f_v}^-}}\;\;
\prod_{e_v}\left[\iota_e^{A_{e1}\cdots A_{eF}}\;\prod_{f\in\partial v}C^{m_{e_f}^+m_{e_f}^-}_{A_{e_f}}\right]\\
&\prod_{(e,f)\in\partial v}\left[\epsilon^{n_{e_f}^+n_{e'_f}^+}\;\;\epsilon^{n_{e_f}^- n_{e'_f}^-}\;\;R^{j_f^+}_{m_{e_f}^+ n_{e_f}^+}(g_{e_f}^+)\; \;R^{j_f^+}_{m_{e_f}^- n_{e_f}^-}(g_{e_f}^-)\right]
\end{split}
\eeq
with $j^{\pm}\equiv\frac{|\gamma\pm1|}{2}j$. Furthermore, $R^j(g)$ denotes a Wigner matrix, $ C^{m_e^+ m_E^-}_{A_e}$ a Clebsch-Gordan coefficient and $\epsilon^{n_e^+ n_e^-}$ represent the unique two-valent intertwiners of $\group{SU}{2}$ (see \cite{Ding:2010fw}). \\
It follows immediately that 
\beq
\label{eqn:vertamp}
\begin{split}
\mathcal{A}^E_v(g_f)&=\sum_{j_f,\iota_e}\;\;\scal{T^E_{\Gamma_v,j_f,\iota_e}|\mathcal{A}_v}{}\;T^E_{\Gamma_v,j_f,\iota_e}(g_f)\\
&:=\sum_{j_f,\iota_e}\left(\prod_{f\in\partial v}\sqrt{d_{j_f^+}\;d_{j_f^-}}\right)  \mathcal{A}^E_v(j_f,\iota_e)\;T^E_{\Gamma_v,j_f,\iota_e}(g_f)
\end{split}
\eeq
defines the EPRL vertex amplitude with 
\beq
\label{eqn:vertexTr}
A_v^E(j_f, \iota_e)=\sum_{\iota_e^+,\iota_e^-}
\mathrm{Tr}_v\left(\bigotimes_{e_v}(\iota^+_{e_v}\otimes\iota^-_{e_v})^{\dagger}\right)
\prod_{e_v} f^{\iota_{e_v}}_{\iota_{e_v}^+,\iota_{e_v}^-}
\eeq
where $ f^{\iota_e}_{\iota_e^+,\iota_e^-}$ are the well known fusion coefficients \cite{Engle:2007uq}. At last one has to replace \eqref{eqn:vertamp} in \eqref{eqn:zbf} to obtain the full transition amplitude of the EPRL model. Expanding the delta function in \eqref{eqn:zbf} in terms of spin-network function and integrating over the group elements gives
\beq
\label{eqn:zeprla}
 Z[\kappa]=\sum_{j_f,\iota_e}\;\prod_f\;d_{j_f^+}\;d_{j_f^-}\;\prod_v\;\mathcal{A}^E_v(j_f,\iota_e)\;.
 \eeq
 \mbox{}\\
{\bf Remark}\\
In order to evaluate the fusion coefficients $ f^{\iota_e}_{\iota_e^+,\iota_e^-}$ by graphical calculus it is convenient to work with $3j$-symbols instead of Clebsch-Gordan coefficients. When replacing the Clebsch-Gordan coefficients we have to multiply by an overall factor $\prod_e\prod_{f_e}\sqrt{2j_{f_e}+1}$.

\subsection{Spin-foam projector}
Instead of using spin-foams as a tool to compute 'transition' amplitudes between spin-networks one is tempted to interpret spin-foams as a projector onto the physical Hilbert space. Given any couple of ingoing and outgoing kinematical states $\psi_{out}$, $\psi_{in}$,
the Physical scalar product can be formally defined by 
\beq
\langle \psi_{out}| \psi_{in}\rangle_{\text{phys}}:= [\eta(\psi_{out})](\psi_{in})
\eeq
where $\eta$ is a projector (Rigging map) onto the Kernel of the Hamiltonian constraint.
Suppose that the transition amplitude $Z$ 
\beq
\langle \psi_{out}|Z|\psi_{in}\rangle:=[\eta(\psi_{out})](\psi_{in})
\eeq
 can be expressed in terms of a sum of spin-foams $(\kappa,\rho,\iota)$ with boundary $\partial (\kappa,\rho,\iota)=\psi_{out} \bigcup \psi_{in} $. To realize that, we first have to reconsider \eqref{eqn:zbf} for a foam $\kappa$ with non-empty boundary $\partial\kappa\neq\emptyset$. Then\footnote{If we would also integrate over group elements in the boundary then $Z^{BF}=\int\delta(F)$ would become singular.} 
\beq
\label{eqn:zbound}
Z[\kappa]=\int_{\group{SO}{4}^{\mathcal{V}_{int}}}\intd{g_{f_v}} \prod_{f\in\kappa}\delta\left(\prod_{v\in\partial f}g_{f_v}g_l\right)\prod_{v\in\mathcal{V}_{int}} \mathcal{A}_v(g_{fv})
\eeq
where 
\beq
g_l=\begin{cases} h_l & \text{if}\quad f\cap\partial\kappa=l\\1&\text{otherwise}\end{cases}\quad.
\eeq
Equation \eqref{eqn:zbound} can be interpreted as a function on the boundary graph $\partial\kappa$. That is to say
\beq
\label{eqn:zbound1}
\begin{split}
Z[\kappa]= &\sum_{j_f,\iota_e}\;\prod_f\;d_{j_f^+}\;d_{j_f^-}\;\prod_{v\in\mathcal{V}_{int}}\;\mathcal{A}^E_v(j_f,\iota_e)\\
&\times\sum_{j_l,\iota_n} \left(\prod_{l\in\partial\kappa^{(1)}}\frac{1}{\sqrt{d_{j_l^+}\;d_{j_l^-}}}\right)
T^E_{\partial\kappa,j_l,\iota_n}(h_l)
\end{split}
\eeq
in the EPRL sector. Here,  $\partial\kappa^{(1)}$ is the set of boundary links. Unfortunately, \eqref{eqn:zbound1} defines an $\group{SO}{4}$ spin-network function while the kinematical Hilbert space of the canonical theory is spanned by $\group{SU}{2}$ functions. It is, however, easy to resolve that problem: when restricting the boundary elements $h_l\in\group{SU}{2}\subset\group{SO}{4}$ then $T^E_{\partial\kappa,j_l,\iota_n}(h_l)$ is a true $\group{SU}{2}$ spin-network function. Indeed
\beq
T^E_{\partial\kappa,j_l,\iota_n}(h_l)=
\left(\prod_{l\in\partial\kappa}\sqrt\frac{d_{j_l^+}\;d_{j_l^-}}{d_{j_l}}\right) \ket{S(\partial\kappa,j,\iota})_N
\eeq
where $\ket{S}_N$ is a normalized spin-network function on $\group{SU}{2}$ (see \ref{snormalization}). This finally implies
\beq
\label{eqn:scalZ1}
\scal{\psi_{out}|Z|\psi_{in}}{}=\sum_{j_f,\iota_e}\;\prod_f\;d_{j_f^+}\;d_{j_f^-}\;\prod_{l\in\partial\kappa^{(1)}}\frac{1}{\sqrt{d_{j_l}}}\;\prod_{v\in\mathcal{V}_{int}}\;\mathcal{A}^E_v(j_f,\iota_e) \;.
\eeq
In the next section we will compute an easy example of such an amplitude.
\section{New solutions to the Euclidean Hamiltonian constraint}
In the following we compute new solutions to the Euclidean Hamiltonian constraint by employing spin-foam methods. We show that
\beq
\label{eqn:physscal1}
\sum_{\phi}\langle \psi_{out} | Z[\kappa] | \phi \rangle\langle \phi|\hat{H}^{(m)}| \psi_{in}\rangle=0\,.
\eeq
in the Euclidean sector with $\gamma=1$ and $s=1$, where $\kappa$ is an easy 2-complex with only one internal vertex.
 \subsection{Trivalent nodes}
 \label{ssec:trivalent}
Consider the simplest possible case given by an initial and final state $\ket{\Theta}$, characterized by two  trivalent nodes joined by three links:
\beq
|\Theta(j_i,j_j,j_k)\rangle=\Bigg|\left.
\makeSymbol{\includegraphics{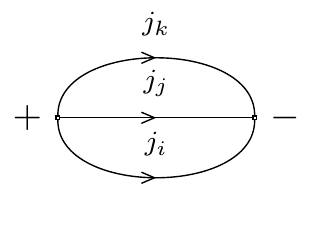}}			
 \right\rangle
 \label{tetha}
\eeq
As shown in chapter \ref{sec:Ham} the only states produced by the Hamiltonian $\hat{H}^{(m)}$ acting on a node, are given by a linear combination of spin-networks that differ from the original one by the presence of an extraordinary link. In particular the term $\langle s|\hat{H}^{(m)}| \Theta(j_i,j_j,j_k)\rangle$,  will be non vanishing only if $\langle s |$ is of the kind:
\beq
\langle s|=\left <			
\makeSymbol{\raisebox{.1\height}{\includegraphics{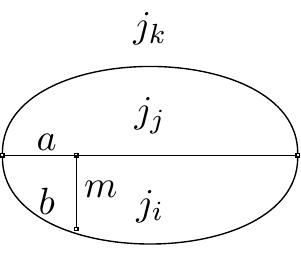}}}\right.
 \Bigg |
 \label{s}
\eeq

\begin{figure}[t]
 \begin{center}
	\includegraphics{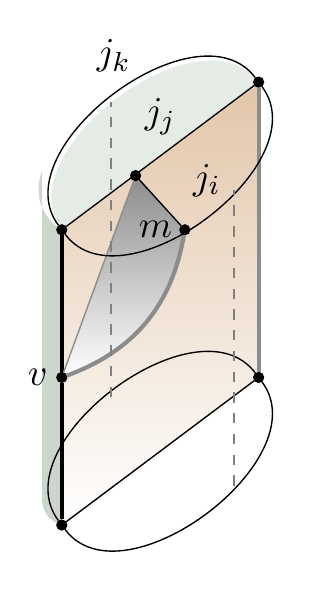}
 \end{center}
\caption{Two-complex $\kappa$ with on internal vertex.}
\label{fig:foam}
\end{figure}
\begin{figure}[t]
 \begin{center}
	\includegraphics{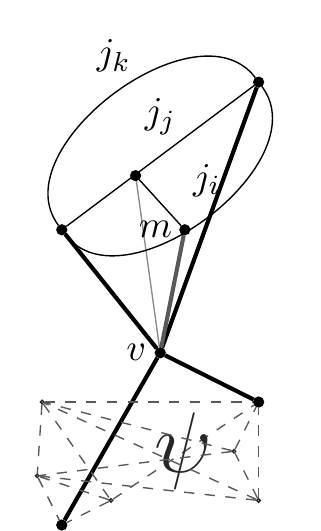}
 \end{center}
\caption{Two-complex $\kappa'$ with on internal vertex and arbitrary $\psi$.}
\label{fig:foam2}
\end{figure}

The simplest two-complex $\kappa(\Theta,s)$ with only one internal vertex defining a cobordism between $\ket{\Theta}$ and $\ket{s}$ is a tube $\Theta\times[0,1]$ with an additional face between the internal vertex and the new link $m$ (see Fig. \ref{fig:foam}). \\
The computation of \eqref{eqn:zbound1} for $\kappa(\Theta,s)$ is straightforward when using graphical calculus. Since the space of three-valent intertwiners is one-dimensional and all labelings $j_f$ are fixed by the states $\ket{s},\ket{\Theta}$ the first sum in \eqref{eqn:zbound1} is trivial. Thus 
\beq
\scal{\Theta|Z[\kappa]|s}{}:=W_{E}(\kappa,\Theta,s)
=\mathcal{A}_f\;\mathcal{B}\;\mathcal{A}^E_v(j_f,\iota_e) 
\eeq 
where $\mathcal{A}_f=\prod_f\;d_{j_f^+}\;d_{j_f^-}$ are the face amplitudes and $\mathcal{B}=\prod_{l\in\partial\kappa^{(1)}}\frac{1}{\sqrt{d_{j_l}}}$ are the boundary amplitudes. The evaluation of the trace in $A_v^E$ is equivalent to evaluating the boundary spin-network $\Gamma_v$ of the vertex $v$ \cite{Ding:2010fw} at $1$. The reader can easily convince herself that $\Gamma_v=s$ and therefore with \eqref{eqn:basicrule2} gives
\beq
Tr_v(\bigotimes_e\iota_e^+\iota_e^-)=(-)^{j_i^+ +j_j^+-j_k^+}(-)^{j_i^-+j_j^- -j_k^-}\sixj{j_i^+}{b^+}{j_j^+}{a^+}{ j_k^+}{m^+}\sixj{j_i^-}{b^-}{j_j^-}{a^-}{ j_k^-}{m^-}\;.
\eeq
where the sign factor is due to the orientation of $s$ (see \eqref{trace_part}).The fusion coefficients contribute four $9_j$ symbols since
\beq
\begin{split}
f^{\iota_e}_{\iota_e^+,\iota_e^-}=\sqrt{d_a d_b d_c}\;\Bigg[\!\!
\makeSymbol{\includegraphics{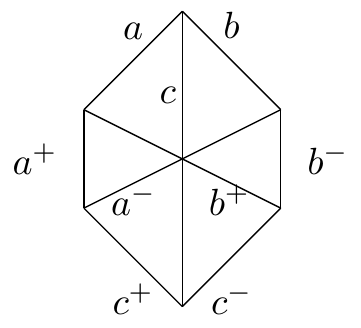}}\!\!\Bigg]
=\sqrt{d_a d_b d_c}\ninej{a}{b}{c}{a^+}{b^+}{c^+}{a^-}{b^-}{c^-}
\end{split}
\eeq	
where the dimension factors come from the replacement of Clebsch-Gordan coefficients by $3j$-symbols (see Sec.\ref{ssec:model}). The full amplitude is
\beq
\begin{split}
W_E (\kappa,\Theta,s)=&\mathcal{A}_f\mathcal{A}_e\mathcal{B} (-)^{j_i^+ +j_j^+-j_k^+}(-)^{j_i^-+j_j^- -j_k^-}
\;\;\sixj{j_i^+}{b^+}{j_j^+}{a^+}{ j_k^+}{m^+}\sixj{j_i^-}{b^-}{j_j^-}{a^-}{ j_k^-}{m^-}
 \\
& \times
\begin{Bmatrix}  j_i & j_j  & j_k \\ j_i^+ & j_j^+  & j_k^+ \\ j_i^- & j_j^-  & j_k^- 
\end{Bmatrix}
\begin{Bmatrix}  j_i & a  & m \\ j_i^+ & a^+  & m^+ \\ j_i^- & a^-  & m^- 
\end{Bmatrix}
\begin{Bmatrix}  j_j & b  & m \\ j_j^+ & b^+  & m^+ \\ j_j^- & b^-  & m^- 
\end{Bmatrix}
\begin{Bmatrix}  a & b  & j_k \\ a^+ & b^+  & j_k^+ \\ a^- & b^-  & j_k^- 
\end{Bmatrix}
\end{split}
\label{Wj}
\eeq
with $\mathcal{A}_e=d_{j_i}\,d_{j_j}\,d_{j_k}\,d_{a}\,d_{b}\,d_{m}$. Let us fix $\gamma=1$ then $j^+=j$ and $j^-=0$ and \eqref{Wj} reduces to
\beq
W_E (\kappa,\Theta,s)|_{\gamma=1}=(d_a\, d_b\, d_m)^{1/2}(-)^{j_i+j_j-j_k}\sixj{j_i}{b}{j_j}{a}{ j_k}{m}
\label{eqn:Wj1}
\eeq
where we have used
\beq
\ninej{a}{b}{c}{a}{b}{c}{0}{0}{0}=\frac{1}{\sqrt{d_a\,d_b\,d_c}}\;.
\eeq
With the previous results and \eqref{amplitude} we are now able to compute \eqref{eqn:physscal1}. Note, that in \eqref{trace_part} the new created links labeled by $a,b,m$ are not normalized but the spin-foam amplitude has been constructed such that $\ket{s}$ is normalized. Taking the scalar product $\scal{s|H|\Theta}{}$ gives, therefore an additional factor $\frac{1}{\sqrt{d_a\,d_b\,d_c}}$. This yields \footnote{with Lapse function $N_{\vgraph}=1$ } 
\beq
\begin{split}
&
\sum_{s} W_E(\kappa,s,\Theta)|_{\gamma=1}\scal{s|\hat{H}^{(m)}|\Theta}{}
 \\
& =\sum_{a,b} 
\begin{Bmatrix}  j_i & j_j  & j_k \\  b & a & m
\end{Bmatrix}
\sum_c\; \lambda^{mj_i}_a 
        \lambda^{mj_j}_b \lambda^{mj_k}_c\;d_a d_b d_c\!\sum_{\beta(j_i,j_j,m,c)}  \!\!\!\! V{}_{j_k}{}^\beta (j_i,j_j,m,c) \\
&\times\;\left[ \lambda_c^{m\beta}(-)^{a+j_j+c}\; \tinysixj{a}{\beta}{j_j}{m}{c}{j_i}\tinysixj{a}{m}{b}{c}{j_k}{j_j}                 
  - 
   \lambda^{mj_k}_c (-)^{b+j_i+c}\tinysixj{j_i}{m}{b}{\beta}{c}{j_j}\tinysixj{a}{c}{b}{m}{j_k}{j_i}
   \right]  \\
   &+[j_k\leftrightarrow j_i]+[j_k\leftrightarrow j_j]\\
 \end{split}
 \label{BFm}
\eeq
The last two terms are equivalent to the first term when exchanging $j_k\leftrightarrow j_j$ respectively $j_i$ and correspond to the other extraordinary links. In fact the EPRL spin-foam reduces just to the SU(2) BF amplitude that is just the single $6j$ left in the first line.
Now using the  definition of a $9j$ in terms of three $6j$'s \eqref{9jdef}
equation \eqref{BFm} becomes
\beq
\begin{split}
&\sum_{s} W_E(\kappa,s,\Theta)|_{\gamma=1}\scal{s|\hat{H}^{(m)}|\Theta}{}=\sum_c d_c\!\!\sum_{\beta(j_i,j_j,m,c)} \! \!\! V{}_{j_k}{}^\beta (j_i,j_j,m,c) \\
&\qquad\times \left[
  \sum_b d_b  \lambda^{mj_j}_b (-)^{m+j_i+j_j+c}\; \lambda^{mj_k}_c \lambda^{m\beta}_c \; \ninej{j_i}{j_j}{\beta}{j_k}{m}{c}{j_j}{b}{m}\right.
 -\\
 &
\qquad\qquad\qquad \left.
 \sum_a d_a \lambda^{mj_i}_a (-1)^{m+j_j+j_i+c}
 \ninej{m}{c}{\beta}{a}{m}{j_i}{j_i}{j_k}{j_j}
 \right]
 \\
      &\qquad\qquad\qquad+[j_k\leftrightarrow j_i]+[j_k\leftrightarrow j_j]\\
 \end{split}
 \label{BFm2}
\eeq
The 9j's involved in this expression can be reordered using the permutation symmetries \eqref{9jrow} and \eqref{9jcolumn} giving
\beq
\begin{split}
 &\sum_c d_c \!\!\sum_{\beta(j_i,j_j,m,c)} \! \!\! V{}_{j_k}{}^\beta (j_i,j_j,m,c) \\
&\times
 \left[
  \; \sum_b d_b (-)^{2\beta} \; \ninej{j_i}{\beta}{j_j}{j_k}{c}{m}{j_j}{m}{b}
 -
 \sum_a d_a (-)^{j_k+\beta}
 \ninej{j_j}{j_k}{j_i}{\beta}{c}{m}{j_i}{m}{a}
 \right]
\\
 &  =\sum_c d_c (-1)^{2j_k} \sum_{\beta(j_i,j_j,m,c)}  \!\! V{}_{j_k}{}^\beta (j_i,j_j,m,c) \left[
    \frac{\delta_{\beta j_k}}{d_{j_k}}
 -
\frac{\delta_{\beta j_k}}{d_{j_k}}
 \right] =0
 \\
 \end{split}
 \label{BFm4}
\eeq
In the last expression we have used the summation identity \eqref{9jlastsum}. Equation \eqref{BFm4} shows that the states $\ket{s}_{phys}=\sum_{s}W_E(\kappa, \Theta,s)|_{\gamma=1}\bra{s}$ are solutions of the (Euclidean) Hamiltonian constraint if $s$ is of the form \eqref{s}. However, each term depending on one of the three graphs which differ by its extraordinary link vanishes separately. This  suggest that the solution we have constructed is very likely not the most arbitrary solution for trivalent nodes.\\
\mbox{}\\
{\bf Remarks}
\begin{itemize}
\item {\emph{The role of the volume}}\\
It is noteworthy that  the spin-foam amplitude selects only those terms which depend on the diagonal elements on the volume. The consequences of this behavior are manifold.\\
 First, it simplifies the calculation since we do not have to evaluate the volume explicitly. If $m=1/2$ this would not be a problem since then the volume is already diagonal and can be computed easily \cite{Brunnemann:2004xi}. But if $m\neq 1/2$ or in higher valent cases the structure of the volume operator is very complicated and is the major obstacle for computing solutions of the Hamiltonian. Indeed we show in the next section that the above property carries over to higher-valent nodes and therefore enables to compute more solutions.\\
On the other side this behavior supports the conjecture that the states constructed are not the most arbitrary solutions but only a special class. Looking more closely at the spin-foam amplitude this is hardly surprising. When setting the Barbero-Immirzi parameter $\gamma=1$ we restrict to $BF$-theory (in the spin-foam framework). The Hamiltonian of $BF$-theory is essentially given by the curvature $F$ and the only part of $\hat{H}^{(m)}$ influencing the spin-network structure of $\ket{s}_{phys}$ is again the curvature; the volume just yields an overall factor. This shows to some extend the consistency between the models.
\item {\emph{Arbitrary cobordism}}\\
The result \eqref{BFm4} is obviously not sensitive to the orientations of $\Theta$ respectively $s$ since a change in the orientation would give the same sign factor in $A^{(m)}(j_i,a|j_j,b|j_k)$ as in $W_E(\kappa,\Theta,s)$. The crucial ingredient of $W_E(\kappa,\Theta,s)$ is the appearance of the $6j$-symbol (see \eqref{BFm} and \eqref{BFm2}). Thus, \eqref{eqn:physscal1} also vanishes if we consider a more general complex $\kappa'$ as long as $W_E(\kappa',\psi,s)$ still depends on the same $6j$-symbol and the rest does not depend on the spins $a,b$. For example we could work with a cobordism between an arbitrary state $\psi$ and $s$ such that all faces of $\kappa$ wind up in the same internal vertex (see Fig. \ref{fig:foam2}).
\end{itemize}

\subsection{Four valent nodes}
Let us now turn to the case with $\psi_{in}=\psi_{out}=\ket{\vgraph_4}$ where
\beq
\ket{\vgraph_4}=\Bigg|\left.
\makeSymbol{\includegraphics{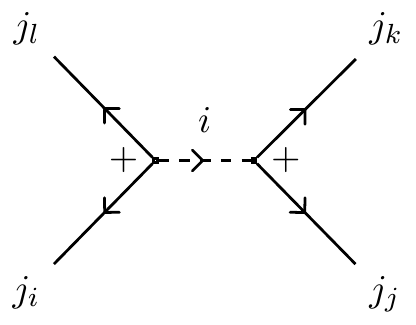}}\right\rangle\quad.
\eeq
The matrix element $\scal{s|\hat{H}^{(m)}|\vgraph_4}{}$ is non-vanishing iff $\ket{s}$ is of the form
\beq
\ket{s}=\Bigg|\left.\makeSymbol{\includegraphics{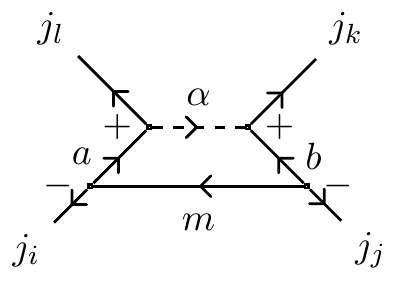}}\right\rangle\quad.
\eeq
We choose again an easy complex $\kappa$ of the form Fig.\ref{fig:foam} with one additional face $j_l$. The vertex trace in \eqref{eqn:vertexTr} can be evaluated by graphical calculus
\beq
\begin{split}
&\mathrm{Tr}_v(\bigotimes_{e_v}\iota^{\pm}_e)=
\makeSymbol{\includegraphics{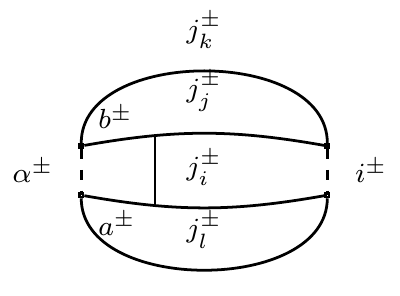}}\\
&=(-)^{b^{\pm}-a^{\pm}+\alpha^{\pm}+m^{\pm}+j_l^{\pm}-j_j^{\pm}}\; d_{i^{\pm}}\;\tinysixj{\alpha^{\pm}}{j_i^{\pm}}{i^{\pm}}{a^{\pm}}{m^{\pm}}{j_l^{\pm}}
\tinysixj{\alpha^{\pm}}{j_j^{\pm}}{i^{\pm}}{a^{\pm}}{m^{\pm}}{j_k^{\pm}}	
\end{split}			
\eeq
The fusion coefficients $f^{\iota_e}_{\iota^+_e\iota^-_e}$ give two $9j$ symbols for the two trivalent edges and two $15j$- symbols for the two four-valent edges. As in the above section the fusion coefficients reduce to $1$ when setting $\gamma=1$. When taking the scalar product \eqref{eqn:scalZ1} the internal links labeling the intertwiner can be in principle treated like the real links and we obtain 
\beq
W_E(\kappa,\vgraph_4,s)
=  \sqrt{d_a\;d_b\;d_m} \;(-)^{b-a+\alpha+m+j_l-j_j} \;\tinysixj{\alpha}{j_i}{x}{a}{m}{j_l}\tinysixj{\alpha}{j_j}{x}{b}{m}{j_k}\;
\eeq
Taking the scalar product with the Hamiltonian as in \eqref{BFm} yields
\beq
\begin{split}
&\sum_{s}W_E(\kappa,\vgraph_4,s)\scal{s|\hat{H}^{(m)}|\vgraph_4}{}=\sum_{a,b,c} d_a d_b d_c \sum_{\alpha}\ d_{\alpha}\;(-)^{b-\alpha}\;  \tinysixj{\alpha}{j_i}{i}{a}{m}{j_l}\tinysixj{\alpha}{j_j}{i}{b}{m}{j_k}\sum_{\beta,\gamma} V_{i,j_k}^{\;\;\gamma, \beta}\\
&\times
  \Big[
  (-1)^{\alpha+\beta-c-j_j}
\tinysixj   { \gamma}  { a} {\alpha}    {j_i} {m}  {j_l} 
\tinysixj    {\gamma}  {c} {\alpha} {\beta} {m}  {j_j} 
\tinysixj     {\alpha} {m}   {j_j}{j_k} {c}  {b} 
-(-1)^{c+\gamma-j_k-b} \tinysixj{ \gamma}   { m}  {j_j} {c} {\beta} {b} 
\tinysixj{  \gamma}   { a}  {\alpha} {j_i} {m} {j_l} \tinysixj{ \gamma}   { j_k}  {\alpha} {c} {m} {b}
\Big]
\end{split}
\label{sp}
\eeq
where we have used $(-1)^{2a+2j_l+2b}=1$. Summing over $a$ and using the orthogonality relation \eqref{orthogonality} and $(-1)^{2j_i+2a+2m}=1$ gives
\beq
\begin{split}
&\sum_{s}W_E(\kappa,\vgraph_4,s)\scal{s|\hat{H}^{(m)}|\vgraph_4}{}
=\sum_{b,c}  d_b d_c \sum_{\alpha} d_{\alpha} \sum_{\beta,\gamma} V_{i,j_k}^{\;\;\gamma, \beta} \delta_{i,\gamma}\\
&\times 
  \Big[
  (-1)^{\beta-b-c-j_j-2m} \tinysixj{\alpha}{j_j}{x}{b}{m}{j_k}
\tinysixj    {\gamma}  {c} {\alpha} {\beta} {m}  {j_j} 
\tinysixj     {\alpha} {m}   {j_j}{j_k} {c}  {b} 
- (-1)^{c+i+\alpha-j_k+2m} \tinysixj{\alpha}{j_j}{x}{b}{m}{j_k} \tinysixj{  \gamma}   { m}  {j_j} {c} {\beta} {b}   \tinysixj{  \gamma}   { j_k}  {\alpha} {c} {m} {b}
\Big]
\end{split}
\label{sp1}
\eeq
Note, the three 6j's in the two terms of \eqref{sp} define a 9j summing over the indexes $\alpha$ and $b$ respectively:
\beq
\begin{split}
&\sum_c d_c \sum_{\beta} V_{i,j_k}^{\;\;i, \beta}\;
  \Big[\sum_b d_b
  (-1)^{b+\beta-c-j_j+2b+2m} \tinyninej{m}{c}{\beta}{b}{m}{j_j}{j_j}{j_k}{i}
- \sum_{\alpha} d_{\alpha}(-1)^{c+i-j_k+\alpha+2m} \tinyninej{i}{\alpha}{m}{j_j}{i}{j_k}{\beta}{m}{c}\Big]\\
&=\sum_c d_c  \sum_{\beta} V_{i,j_k}^{\;\;i, \beta}\;\;
  \Big[\sum_b d_b
  (-1)^{2\beta+j_k+\gamma+j_j} \tinyninej{\gamma}{j_k}{j_j}{\beta}{c}{m}{j_j}{m}{b}
- \sum_{\alpha} d_{\alpha}(-1)^{2j_k+\beta+\gamma+j_j} \tinyninej{j_j}{j_k}{\gamma}{\beta}{c}{m}{\gamma}{m}{\alpha}\Big]
\end{split}
\label{sp3}
\eeq
In the second line we used the permutation symmetry (see the Appendix). With \eqref{9jlastsum} we obtain the final result
\beq
\begin{split}
\sum_c d_c  \sum_{\beta} V_{i,j_k}^{\;\;i, \beta}\;\;   (-1)^{3j_k+\gamma+j_j}
  \Big[\frac{\delta_{\beta j_k}}{d_{j_k}}- \frac{\delta_{\beta j_k}}{d_{j_k}}\Big]=0
\end{split}
\label{sp5}
\eeq
As for the trivalent vertex the spin-foam amplitude just takes those elements into account which depend on the diagonal Volume elements. In contrast to the trivalent case this partly depends on the choice of $\Psi_{out}=\ket{\vgraph_4}$. If one chooses for example an other four valent vertex $\Psi_{out}=\ket{\vgraph'_4(x)}$ and $\Psi_{in}=\ket{\vgraph_4(i)}$ where $x$ respectively $i$ label the intertwiner then one obtains $\delta{\gamma,x}$ in \eqref{sp1}. However, the part of the volume depending on the links where the curvature acts is still diagonal, namely $\delta_{\beta,j_k}$ remains unchanged in \eqref{sp5}. This shows that the above calculation can be easily extended to n-valent nodes since the curvature always acts locally on three links while the influence of the volume on the rest of the internal links is unimportant. 
\section{Conclusions}
LQG is grounded on two parallel constructions; the canonical and the covariant ones. One of the bigger missing theoretical ingredients of this road to quantum gravity is the relation between these two. In this paper we were mainly concerned with the following questions: The EPRL-FK with the KKL extension shares the same kinematics of LQG, do they share also the same \emph{dynamics?} Can we really use the EPRL-FK as defining the Physical Hilbert Space? A first step to find an answer to that important questions is to construct a simple spin-foam amplitude which annihilates the Hamiltonian constraint as argued in \eqref{eqn:physscal1}. Indeed we found that in the euclidean sector with signature $s=1$ and Barbero-Immirzi parameter $\gamma=1$ the Euclidean Hamiltonian constraint is annihilated by a spin-foam amplitude $Z[\kappa]$ where $\kappa$ is a simple two-complex with only one internal vertex. Even though we considered only a very special case this has some important consequences:\\
First, even neglecting their spin-foam origin, the one vertex amplitudes of BF theory are \emph{new explicit analytic solutions of the Hamiltonian theory} and represent a proper subspace of the Physical Hilbert space.\\
Second the equation \eqref{BFm} vanishes for each triple of edges, this means that the 6j symbol associated to every face is annihilated by the Euclidean scalar constraint. This is a generalization of the work by Bonzom-Freidel in the context of 3d gravity. In \cite{Bonzom:2011hm} they found that the 6j (a physical state in 3d) is annihilated by a suitable quantization of the 3d scalar constraint  $F=0$ rewritten, following \cite{thomas3d}, as $EEF$. Their result holds exactly only for the choice $m=1$ (even if a generalization to higher spin, involving a proper redefinition of the quantum constraint, is discussed, see \cite{Bonzom:2011hm}). Here, we showed that the 6j-symbols one obtains in the one vertex expansion annihilates for \emph{arbitrary spin $m$} the complete non polynomial, density constraint $\frac{EEF}{\sqrt{\det E}}$.\\
It was already pointed out at the end of Sec.\ref{ssec:trivalent} that the spin-foam amplitude diagonalizes the Volume. As we have seen on the one hand this behavior proves to be very useful when computing \eqref{eqn:physscal1} for higher valent nodes. On the other hand this indicates that the solutions are not the most arbitrary solutions but are closely related to BF-solutions, which is not surprising since setting $\gamma=1$ in the spin-foam model yields BF-theory. But classically the flat solutions are \emph{not} the only solutions for the Euclidean theory.\\
Of course many open questions remain, e.g.:
\begin{itemize}
\item The general case $\gamma\neq1$ and the Lorentzian signature model: in this case there are indications (work in progress) that seem to suggest the use of projected spin-networks \cite{Alexandrov:2002br}.

\item The relation between the  geometric structure of the $n-j$'s involved in the amplitude, the Hamiltonian and the simplicial geometry \cite{Dittrich:2011ke} deserves also to be investigated for example along the lines of \cite{Bonzom:2011tf}.

\item The relation with other regularization \cite{EmanueleReg} and quantization programs, e.g. \cite{Giesel:2006uj} for the canonical theory and the EPRL-FK  can be analysed along the same lines described here.
\end{itemize}
\mbox{}\\
\mbox{}\\
{\bf{Acknowledgements}}\\
AZ wants to thank ``Universit\"at Bayern e.V.'' for financial support. EA wishes to thank V.Bonzom and L.Freidel for useful discussions and a clarification of their construction, during a visit to Perimeter Institute.
\appendix
\section{Graphical Calculus}
\label{graphical}
In order to compute the matrix elements of the Hamiltonian constraint operator as well as the vertex amplitudes in the spin-foam model one has to make extensive use of recoupling identities for $\group{SU}{2}$. In this context graphical calculus can be very useful to keep track of indices and sign. In this appendix we summarize the graphical methods used in the main text.
\subsection{Basic Elements}
Our convention is applicable in pure recoupling  theory as well as in computations involving group elements. The convention is mainly based on \cite{BrinkSatchler68} including some minor improvements.\\
\begin{itemize}
\item\emph{Irreducible Representation:} Multiplication with an orthonormal vector in the $j$-representation of $\group{SU}{2}$ is represented by
\begin{align}
\begin{split}
\makeSymbol{\raisebox{.75\height}{\includegraphics{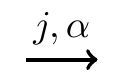}}}&=\ket{j,\alpha}\\
\makeSymbol{\raisebox{.75\height}{\includegraphics{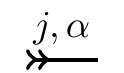}}}&=\bra{j,\alpha}\,,
\end{split}
\end{align}
where the italic letters $j\in\frac{1}{2}\mathbb{N}$ label the irreducible representation of $\group{SU}{2}$ and greek letters $-j\geq \alpha\leq j$ represent  magnetic quantum numbers. To avoid an unnecessary cumulation of labelings we will suppress  the label $j,\alpha$ if there is no danger of confusion.
\item \emph{Wigner-$R$-matrix:} 
\begin{align}
\makeSymbol{\raisebox{.25\height}{\includegraphics{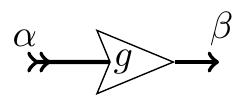}}}
			=R^{\alpha}\,_{\beta}(g)
\end{align}
Note: $\alpha$ transforms in the dual representation while $\beta$ transforms in the standard representation. Thus $\alpha$ must be contracted with an intertwiner $\iota^{\dots}\;_{\dots\alpha\dots}$ while $\beta$ gets contracted by the dual $\iota^{\dots\beta\dots}\:_{\dots}$.
\item\emph{Wigner $3j$-Symbol:}
\begin{align}
\begin{split}
\threej{a}{\alpha}{b}{\beta}{c}{\gamma}=&\makeSymbol{\includegraphics{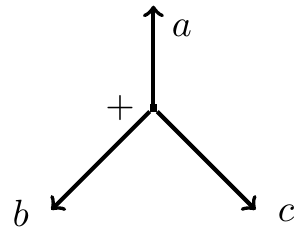}}\\
=(-1)^{a+b+c}&\makeSymbol{\includegraphics{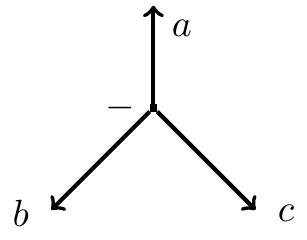}}
\end{split}
\end{align}
The $+$ sign marks anti-clockwise orientation while the $-$ sign marks clockwise orientation.
\item\emph{Dualization:}
\begin{align}
\makeSymbol{\raisebox{.75\height}{\includegraphics{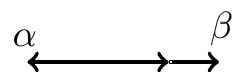}}}
			=\makeSymbol{\raisebox{.75\height}{\includegraphics{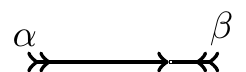}}}
			=\dtensor{j}{\beta}{\alpha}=(-1)^{j+\beta}\delta_{\beta,-\alpha}\\
\makeSymbol{\raisebox{.75\height}{\includegraphics{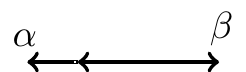}}}
			=\makeSymbol{\raisebox{.75\height}{\includegraphics{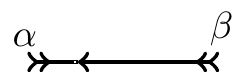}}}
			=\dtensor{j}{\alpha}{\beta}=(-1)^{j-\beta}\delta_{\alpha,-\beta}
\end{align}
This implies:
\begin{align}
\makeSymbol{\raisebox{.75\height}{\includegraphics{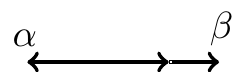}}}
=(-1)^{2j}\makeSymbol{\raisebox{.75\height}{\includegraphics{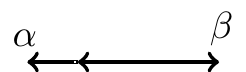}}}
\end{align}
\begin{align}
\begin{split}
\makeSymbol{\raisebox{.75\height}{\includegraphics{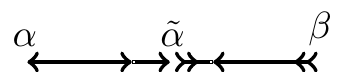}}}&
	=\sum_{\tilde{\alpha}=-j}^j\dtensor{j}{\tilde{\alpha}}{\alpha}\dtensor{j}{\tilde{\alpha}}{\beta}\\
	=\delta_{\alpha,\beta}
	:=&\makeSymbol{\raisebox{.75\height}{\includegraphics{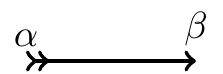}}}
\end{split}
\end{align}
\begin{align}
\makeSymbol{\raisebox{.75\height}{\includegraphics{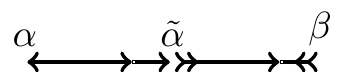}}}
	=(-1)^{2j}\delta_{\alpha,\beta}
\end{align}
Thus,
\begin{align}
\begin{split}
\makeSymbol{\includegraphics{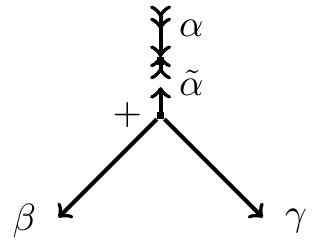}}
=(-1)^{j_a-\alpha}\threej{j_a}{-\alpha}{j_b}{\beta}{j_c}{\gamma}=
\makeSymbol{\includegraphics{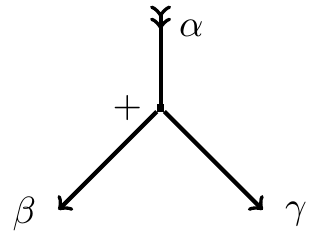}}
\end{split}
\end{align}
and 
\begin{align}
\begin{split}
\label{eqn:invertG}
\makeSymbol{\raisebox{.25\height}{\includegraphics{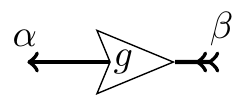}}}&=:\overline{R_\alpha\,^\beta(g)}=\\
\makeSymbol{\raisebox{.25\height}{\includegraphics{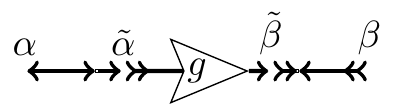}}}
				& =(-1)^{\alpha-\beta}R_{-\alpha}\,^{-\beta}(g)=
\end{split}\\
\makeSymbol{\raisebox{.25\height}{\includegraphics{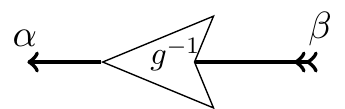}}}
&=R^{\beta}\,_{\alpha}(g^{\scriptscriptstyle{-1}})
\end{align}

\end{itemize}
\mbox{}
\\
\subsection{Recoupling and Simplifications of graphs}
This subsection will give an overview of the basic recoupling identities needed for the evaluation of the Hamiltonian constraint and the spin-foam amplitudes. In the following, we will replace \tikz{\node (a) at (0,0){};\node[draw,inner sep=0] (b) at (0.5,0){} ;\node (c) at (1,0){};\draw[-to,very thick](a)--(b);\draw[-feather,very thick](c)--(b);} by \tikz{\node (a) at (0,0){};\node (b)[draw,inner sep=0] at (0.5,0){} ;\node (c) at (1,0){};\draw[-to,very thick](a)--(b);\draw[very thick](c)--(b);}  on closed lines to simplify the graphs. Furthermore the labels will be suppressed whenever possible. In the following denote the dimension of $j$ by $d_j=2j+1$.
\begin{itemize}
\item\emph{Basic recoupling}
\begin{gather}
\begin{gathered}
\label{eqn:basicrule}
\makeSymbol{\includegraphics{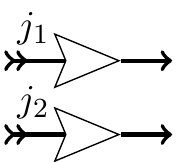}}
	=(R^{j_1}(g))^{\alpha}\,_{\tilde{\alpha}}(R^{j_2}(g))^{\beta}\,_{\tilde{\beta}}\\
	=\sum_{j_3=|j_1-j_2|}^{j_1+j_2}d_{j_3}\sum_{\gamma,\tilde{\gamma}=-j_3}^{j_3}
	\threej{j_1}{\tilde{\alpha}}{j_2}{\tilde{\beta}}{j_3}{\tilde{\gamma}}
	\threej{j_1}{\alpha}{j_2}{\beta}{j_3}{\gamma}\overline{(R^{j_3}(g))^{\gamma}\,_{\tilde{\gamma}}}\\
=\sum_{j_3=|j_1-j_2|}^{j_1+j_2}d_{j_3}
\makeSymbol{\includegraphics{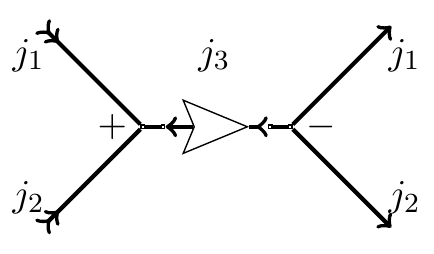}}\\
=(-1)^{2j_3}\sum_{j_3=|j_1-j_2|}^{j_1+j_2}d_{j_3}
\makeSymbol{\includegraphics{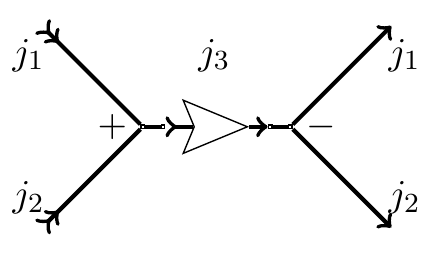}}
\end{gathered}
\end{gather}
Similarly, one finds
\begin{align*}
\makeSymbol{\includegraphics{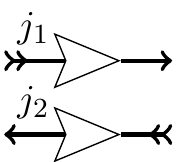}}\;
=(-1)^{2j_2}\sum_{j_3=|j_1-j_2|}^{j_1+j_2}d_{j_3}
\makeSymbol{\includegraphics{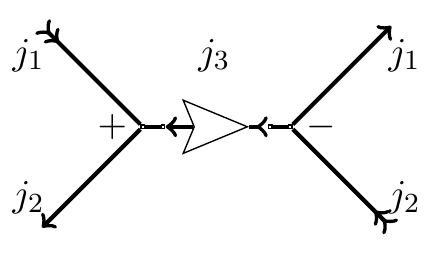}}
\end{align*}

\item\emph{First orthogonality relation}
\begin{align}
\begin{split}
&\sum_{\alpha,\beta}\threej{a}{\alpha}{b}{\beta}{c}{\gamma}\threej{a}{\alpha}{b}{\beta}{\tilde{c}}{\tilde{\gamma}}=\frac{1}{d_c} \delta_{c,\tilde{c}}\delta^{\tilde{\gamma}}\,_{\gamma}\\
	&\makeSymbol{\includegraphics{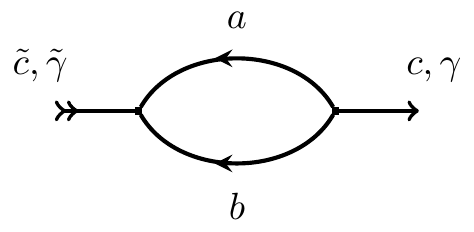}}
=\frac{1}{d_c}
\makeSymbol{\raisebox{.75\height}{\includegraphics{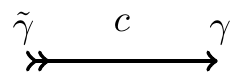}}}
\end{split}
\end{align}

\item\emph{Six-$j$-Symbol}
\begin{align}
\makeSymbol{\includegraphics{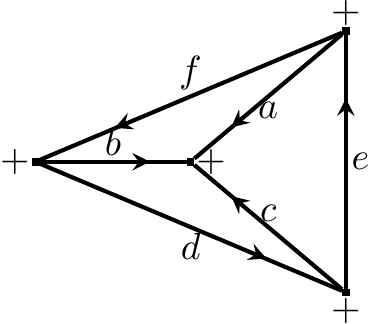}}\quad
		=\quad\sixj{a}{d}{b}{e}{c}{f}
\end{align}

\item{Second orthogonality relation}

\begin{equation}
\sum_{f}
d_m\;d_f 
\left\{\begin{array}{ccc}                      a  & b & f \\                      d & c & e                  \end{array}\right\}
\left\{\begin{array}{ccc}                      a  & c & m \\                      d & b & f                  \end{array}\right\} 
=\delta_{em}
	\label{orthogonality}
\end{equation}

\item{Summation}
The following identity is a crucial ingredient for the computation of the matrix elements of the Hamiltonian operator.
\begin{gather}
\begin{gathered} 
\sum_{\delta,\epsilon,\phi}(-1)^{d+e+f-\delta-\epsilon-\phi}\threej{d}{-\delta}{e}{\epsilon}{c}{\gamma}\threej{e}{-\epsilon}{f}{\phi}{a}{\alpha}\threej{f}{-\phi}{d}{\delta}{b}{\beta}\\
=\sixj{a}{d}{b}{e}{c}{f} \threej{a}{\alpha}{b}{\beta}{c}{\gamma}
\end{gathered}
\end{gather}
Graphically this identity can be encoded in
\begin{align}
\label{eqn:basicrule2}
\makeSymbol{\includegraphics{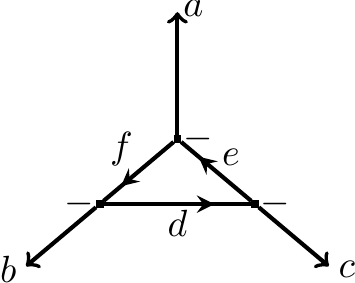}}=
\makeSymbol{\includegraphics{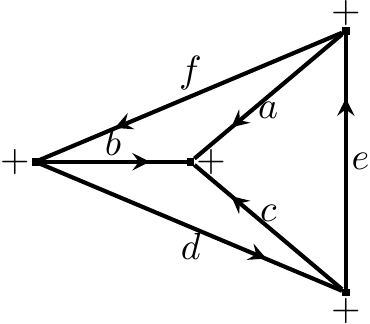}}
\quad\makeSymbol{\includegraphics{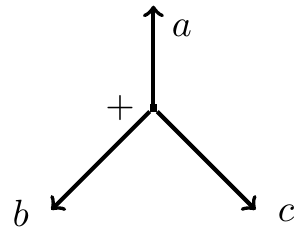}}
\end{align}
\item\emph{Nine-$j$-Symbol}
\\
Definition of a $9j$ Symbol in terms of $6j$'s:
\begin{align}
\sum_x d_x (-1)^{2x} \sixj{a}{c}{b}{d}{x}{p} \sixj{c}{e}{d}{f}{x}{q} \sixj{e}{a}{f}{b}{x}{r}=\ninej{a}{f}{r}{d}{q}{e}{p}{c} {b}
\label{9jdef}
\end{align}
\\
Permutation symmetry:
\\
\begin{enumerate}
\item
\begin{align}
\ninej{j_{11}}{j_{12}}{j_{13}}{j_{21}}{j_{22}}{j_{23}}{j_{31}}{j_{32}} {j_{33}}=\epsilon \ninej{j_{1i}}{j_{1j}}{j_{1k}}{j_{2i}}{j_{2j}}{j_{2k}}{j_{3i}}{j_{3j}} {j_{3k}}
\label{9jcolumn}
\end{align}
\item
\begin{align}
\ninej{j_{11}}{j_{12}}{j_{13}}{j_{21}}{j_{22}}{j_{23}}{j_{31}}{j_{32}} {j_{33}}=\epsilon \ninej{j_{i1}}{j_{i2}}{j_{i3}}{j_{j1}}{j_{j2}}{j_{j3}}{j_{k1}}{j_{k2}} {j_{k3}}
\label{9jrow}
\end{align}
\end{enumerate}
with $\epsilon=1$ for even permutations and $\epsilon=(-1)^R$ with $R=\sum_{ij} j_{ij} $ for odd permutations.

Summation Identity:
\beq
\sum_x d_x\ninej{a}{b}{e}{c}{d}{f}{e}{ f} {x}=\frac{\delta_{bc}}{d_b}\theta(a,b,e)\theta(b,d,f)
\label{9jlastsum}
\eeq
\item\emph{Integration}
The matrix elements of the irreducible representations of $\group{SU}{2}$ provide an orthonormal basis  in the space of square integrable functions of $\group{SU}{2}$ with respect to the Haar measure $\mu(g)$. This can be visualized by the following diagram:
\begin{align}
\label{eqn:Intgroup}
\int\intd{\mu(g)}
\makeSymbol{\includegraphics{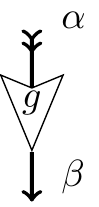}}
\makeSymbol{\includegraphics{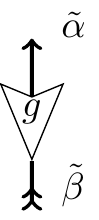}}
=\frac{1}{d_j}
\makeSymbol{\includegraphics{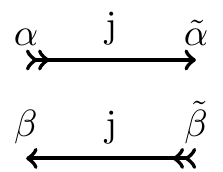}}
\end{align}
The visualization of the integral over higher tensor products works analogously.
\item\emph{Simplify Graphs}
From \eqref{eqn:Intgroup} follow some useful rules for simplifying graphs. In the following completely contracted graphs are symbolized by dashed boxes. 
\begin{align}
\label{eqn:simpgraph1}
\makeSymbol{\includegraphics{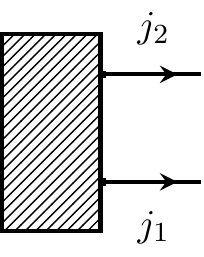}}\quad=\quad
\makeSymbol{\includegraphics{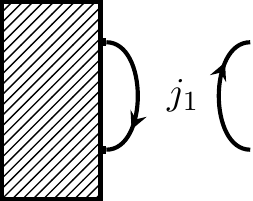}}\qquad\frac{\delta_{j_1,j_2}}{d_{j_1}}	\\
\label{eqn:simpgraph2}
\makeSymbol{\includegraphics{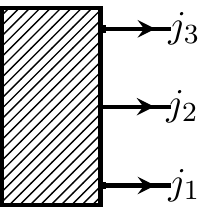}}
		\quad=\quad
\makeSymbol{\includegraphics{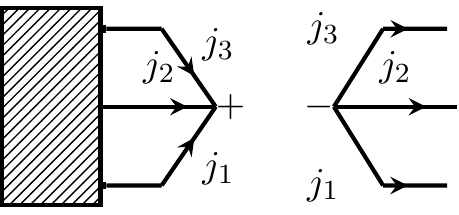}}	
\end{align}
The simplification of graphs with more than three edges can be obtained by applying \eqref{eqn:basicrule} and \eqref{eqn:simpgraph1} respectively \eqref{eqn:simpgraph2}.
\item\emph{Sign-Manipulation}\\
Suppose the three irreducible representations $a,b,c$ obey the Clebsch-Gordan conditions then $(a+b+c)\in\mathbb{N}$ and thus $(-1)^{2a+2b+2c}=1$. This identity is crucial in many calculations in order to simplify the signs or add missing signs of the form $(-1)^{2a}$. Since $a\in\frac{1}{2}\mathbb{N}$ we also have $(-1)^{3a}=(-1)^{-a}$.
\end{itemize}
\section{Grasping}
\label{grasping}
A relevant formula used for the computation of the matrix elements of the Hamiltonian in the 4-valent case can be deduced by the double grasping operators on 4-valent nodes, computed in \cite{scattering3}:
\begin{align}
	&
 \makeSymbol{\includegraphics{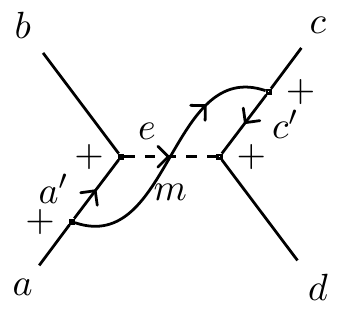}} =\allowdisplaybreaks[4]\nonumber\\
&=
\sum_x 
(-1)^{a'+d+e+x}d_x \sixj{b}{c'}{d}{a'}{x}{e}
 \makeSymbol{\includegraphics{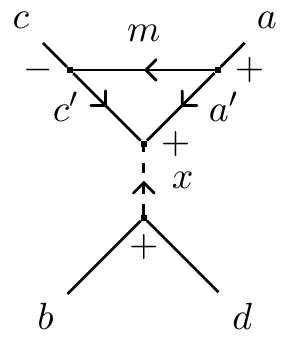}}  \nonumber\allowdisplaybreaks[4]\\
&=
\sum_x 
(-1)^{a+c'+x+m}(-1)^{a'+d+e+x}d_{x}
\sixj{b}{c'}{d}{a'}{x}{e}
\sixj{c'}{a}{a'}{c}{x}{m}
 \makeSymbol{\includegraphics{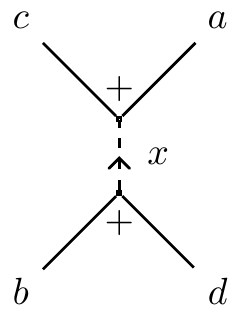}} \nonumber \allowdisplaybreaks[4]\\
&=
\sum_x 
(-1)^{a+c'+x+m}(-1)^{a'+d+e+x}d_{x}\sum_{\alpha}d_{\alpha}(-1)^{a+d+\alpha+x} \cdot \\
&\quad
\cdot
\sixj{b}{c'}{d}{a'}{x}{e}\sixj{c'}{a}{a'}{c}{x}{m}\sixj{b}{c}{a}{d}{\alpha}{x}
 \makeSymbol{\includegraphics{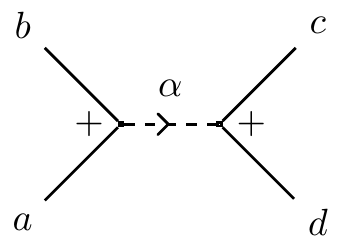}}\nonumber \allowdisplaybreaks[4]\\
&=(-1)^{b+a-c-d}
\sum_{\alpha} d_{\alpha}\sixj{e}{a}{\alpha}{a'}{m}{b}\sixj{e}{c}{\alpha}{c'}{m}{d}
 \makeSymbol{\includegraphics{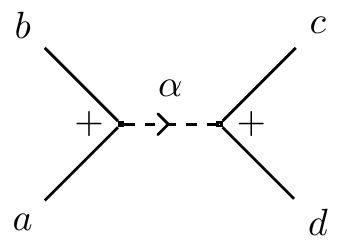}}
\label{ac}
\end{align}
\section{Normalization of the spin-network states}\label{snormalization}
Following \cite{lqgcan2}, we define a spin-network $S=(\Gamma,j_l,i_n)$  as given by a graph $\Gamma$ with a given orientation (or ordering of the links) with $L$ links and $N$ nodes, and by a representation $j_l$ associated to each link and an intertwiner $i_n$ to each node.
As a functional of the connection, a spin-network state is given by 
\begin{equation}
\label{eqn:spinnet}
	\Psi_S[A]=\left\langle A|S\right\rangle\equiv \left(\otimes_l R^{j_l}(h[A,\gamma_l])\right)\llcorner\left(\otimes_n i_n\right)
\end{equation}
where $\llcorner$ indicates the contraction with the intertwiners and $R^{j_l}(h[A,\gamma_l])$ is the $j_l$ representation of the holonomy group element $h[A,\gamma_l]$ along the curve $\gamma_l$ of the gravitation field connection $A$.\\
The scalar product on $\mathcal{H}_{kin}$ is defined via the Ashtekar-Lewandowski measure:
\begin{align}
\begin{split}\label{eqn:scalarnotnorm}
	\left\langle S|S'\right\rangle&=\int\intd{\mu_{AL}} \overline{\Psi_S[A]}\Psi_S[A]\\
		&=\delta_{S,S'} \prod_{e\in \mathcal{E}}\frac{1}{d_{j_e}}\prod_{v\in\mathcal{V}}\mathrm{Tr}(\iota^{\ast}_v\iota_v)
\end{split}
\end{align}
where $\mathcal{E}$ is the set of links and $\mathcal{V}$ is the set of vertices of $\Gamma$. Throughout this paper we use normalized intertwiners such that $\mathrm{Tr}(\iota^{\ast}_v\iota_v)=1$. For a trivalent node this requirement is trivially fulfilled if we use Wigner $3j$-symbols. A higher valent node can be decomposed into trivalent nodes by introducing virtual links. E.g for a 4-valent node we obtain
\beq
\sqrt{d_i}
\makeSymbol{\includegraphics{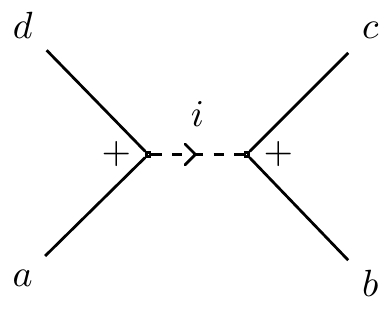}}\eeq
If, additionally, we multiply \eqref{eqn:scalarnotnorm} by 
\beq\nonumber
\prod_{e\in \mathcal{E}}\sqrt{d_{j_e}}
\eeq
we obtain a normalized state $\ket{S}_{N}$ with respect to \eqref{eqn:scalarnotnorm}. 
Consequentially, the recoupling theorem \eqref{ac} applied to normalized spin-network state yields
\begin{equation}
\begin{split}
&\Big|\makeSymbol{\includegraphics{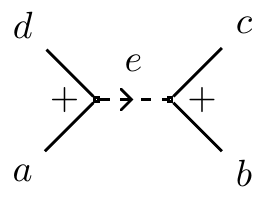}}\Big\rangle_N\\
	&=
\sum_f \sqrt{d_{e}}\sqrt{d_{f}} (-1)^{b+c+e+f}
\quad\sixj{a}{d}{b}{c}{f}{e}\quad
\Big|\makeSymbol{\includegraphics{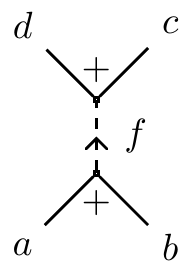}}\Big\rangle_N
\end{split}
\end{equation}
\nocite{*}

\end{document}